\documentclass[10pt,journal,letterpaper,compsoc]{IEEEtran}
\usepackage{longtable}
\usepackage{float} 
\usepackage{subfigure}
\usepackage{hyperref} 
\usepackage{amssymb}
\usepackage[neveradjust]{paralist} 
\usepackage{amsthm}
\usepackage{booktabs}
\usepackage{supertabular}
\usepackage[utf8]{inputenc}

\newtheorem*{theory}{Theory}

\ifCLASSOPTIONcompsoc
\else
\fi
\usepackage[dvips]{graphicx}
\usepackage{array}

\usepackage{eqparbox}
\usepackage{url}

\begin{document}
%
\title{Software Development in Startup Companies: The Greenfield Startup Model}
%
%
%
%

\author{Carmine~Giardino, Nicol\`{o} Paternoster, Michael 
Unterkalmsteiner,~\IEEEmembership{Member,~IEEE,} Tony 
Gorschek,~\IEEEmembership{Member,~IEEE,} and Pekka 
Abrahamsson,~\IEEEmembership{Member,~IEEE}
	
\IEEEcompsocitemizethanks{\IEEEcompsocthanksitem 
C. Giardino is with the Faculty of Computer Science, Free University of 
Bolzano/Bozen, Dominikanerplatz 3, 39100 Bolzano/Bozen, Italy.
}
	
\IEEEcompsocitemizethanks{\IEEEcompsocthanksitem 
N. Paternoster, M. Unterkalmsteiner and T. Gorschek are 
with the Software Engineering Research Lab Sweden, Blekinge Institute of 
Technology, Campus Gräsvik, 371 79 Karlskrona, Sweden.\protect\\
}

\IEEEcompsocitemizethanks{\IEEEcompsocthanksitem 
P. Abrahamsson is with the Department of Computer and Information Science, 
Norwegian University of Science and Technology NTNU, Sem Sælandsvei 7-9,
7491 Trondheim, Norway.}
\thanks{}}

%
%

\markboth{IEEE TRANSACTIONS ON SOFTWARE ENGINEERING,~Vol.~X, No.~X,
XXXX~XXXX}%
{Shell \MakeLowercase{\textit{et al.}}: Bare Demo of IEEEtran.cls for Computer
Society Journals}
%


\IEEEcompsoctitleabstractindextext{%
\begin{abstract}

Software startups are newly created companies with no operating history and
oriented towards producing cutting-edge products. However, despite the 
increasing importance of startups in the economy, few scientific studies 
attempt to address software engineering issues, especially for early-stage 
startups. If anything, startups need engineering practices of the same level or 
better than those of larger companies, as their time and resources are more 
scarce, and one failed project can put them out of business. In this study we 
aim to improve understanding of the software development strategies employed 
by startups. We performed this state-of-practice investigation using a grounded 
theory approach. We packaged the results in the 
Greenfield Startup Model (GSM), which explains the priority of startups to 
release the product as quickly as possible. This strategy allows startups to 
verify product and market fit, and to adjust the product trajectory according 
to early collected user feedback. The need to shorten time-to-market, by 
speeding up the development through low-precision engineering activities, is 
counterbalanced by the need to restructure the product before targeting
further growth. The resulting implications of the GSM outline challenges and 
gaps, pointing out opportunities for future research to develop and validate 
engineering practices in the startup context.

\end{abstract}}


\maketitle

\IEEEdisplaynotcompsoctitleabstractindextext

%
\IEEEpeerreviewmaketitle

\section{Introduction}
\label{intro}
%
%

%
%
%
%
\IEEEPARstart{S}OFTWARE startups launch worldwide every day as a result of an 
increase in new markets, accessible technologies, and venture 
capital~\cite{8491286}. With the term
\textit{software startups} we refer to those organizations focused on
the creation of high-tech and innovative products, with little or no operating
history, aiming to aggressively grow their business in highly scalable markets. 
Being a startup is usually a temporary state, where a 
maturing working history and market domain knowledge leads to the analysis of 
current working practices, thereby decreasing conditions of extreme 
uncertainty~\cite{SMS}.

The research presented in this paper aims at understanding how practitioners 
engineer software development strategies in startups. We focus on the 
structure, planning, and control of software projects, in the period from idea 
conception to the first open beta release. We performed semi-structured, 
in-depth interviews with CEOs and CTOs from 13 startups, covering a wide 
spectrum of themes and iteratively adjusted the developed model according to 
the emerging evidence. With the resulting Greenfield Startup Model (GSM), we 
capture the underlying phenomenon of software development in early-stage 
startups.

New ventures such as \textit{Facebook}, \textit{Linkedin}, \textit{Spotify},
\textit{Pinterest}, \textit{Instagram}, \textit{Groupon} and \textit{Dropbox},
to name a few, are examples of startups that evolved into successful
businesses. Despite many success stories, the vast majority of startups
fail within two years of their creation, primarily due to self-destruction
rather than competition~\cite{Crowne2002}. Operating in a chaotic, rapidly
evolving and uncertain environment, software startups face intense 
time-pressure from the market and are exposed to relentless
competition~\cite{Maccormack2001,Eisenhardt1998}. To succeed in this 
environment startups need to be ready to adapt their product to new market 
demands while being constrained by very limited resources~\cite{Sutton2000}.

From an engineering perspective, software development in startups is challenging
as they work in a context where it is difficult for software processes to 
follow a prescriptive methodology~\cite{Sutton2000, Coleman2005}. Even though 
startups share some characteristics with similar contexts (e.g. small and web 
companies), the combination of different factors makes the specific 
software development context unique~\cite{Blank2005,Sutton2000}. 
Therefore, research is needed to investigate and support the startup 
engineering activities~\cite{Coleman2005}, guide practitioners in taking 
decisions and avoid choices that could easily lead to business 
failure~\cite{Kajko-Mattsson2008}. However, despite the
impressive size of the startup ecosystem~\cite{ISI:000243253000007}, the 
research on software engineering in startups presents a gap~\cite{SMS}. 

With the Greenfield Startup Model (GSM) we aim to contribute to the body of
knowledge on startup software engineering. We created the model as an 
abstraction of reality~\cite{sep-models-science}, based on a systematic 
procedure and grounded on empirical data obtained by the study of 13 cases. 
While the GSM presents the most significant themes in the development 
strategies that characterize these startups' contexts, it does not provide 
guidelines or best practices that should be followed. However, the categories 
in the GSM and the relations among them can provide a common direction, 
vocabulary, and model for future research on software development in startups.

Researchers can use the GSM as a starting point to understand how technical
debt influences the future growth of startup companies. Furthermore, the 
model provides a tool to understand the context in which startups operate, 
which is central when developing 
methods\,/\,models\,/\,tools\,/\,techniques\,/\,practices suited to these types 
of development efforts. Filling gaps on the state-of-practice in startups is 
also beneficial for startup practitioners who can apply the discussed 
strategies to speed up the development initially, although they need also to 
consider the likely drop-down in performance at a later stage. In this regard, 
we identified several commonalities between the issues related to software 
development in startups and the research focused on studying technical 
debt~\cite{Nugroho2011,Izurieta2012}. This paper makes the following 
contributions:
\begin{compactitem}
	\item an empirical investigation into the driving characteristics of 
	early-stage startups
	\item a rigorously developed model that illustrates how and 
	explains why startups perform engineering activities in a certain manner
	\item a discussion on opportunities for future research and potential 
	solutions for the challenges faced by startups
\end{compactitem}

The remainder of this paper is structured as follows. Background and related 
work is covered in Section~\ref{backg}. Section~\ref{resmet}
introduces the research questions and shows the design and execution of the
study. Results are presented in Section~\ref{res:gsm}, illustrating
the GSM. Section~\ref{sect:theory:impl} discusses the most relevant implications
of the GSM. Section~\ref{res:val} compares results of the study to
state-of-the-art in literature. Section~\ref{valt} discusses validity threats. 
The paper concludes in Section~\ref{conc}.

\section{Background} \label{backg}

Looking at the number of new business incubators which appeared in the last
decade one can estimate the importance of startups~\cite{Grimaldi2005}. The wave
of disruption in new technologies has led non-startup companies to be more 
competitive, forcing themselves to undertake radical organizational and 
innovational renewals, in an attempt to behave more like 
startups~\cite{Christensen1997}. However, the implementation of methodologies 
to structure and control development activities in startups is still a 
challenge~\cite{Coleman2008}. Several models have been introduced to drive 
software development activities in startups, however without delivering 
significant benefits~\cite{Coleman2008a,Coleman2008, Sutton2000}.

Software engineering (SE) faces complex and multifaceted obstacles in 
understanding how to manage development processes in the startup context. 
Bach refers to startups as ``a bunch of energetic and committed people without 
defined development processes''~\cite{Bach1998}. Sutton defines startups as 
creative and flexible in nature and reluctant to introduce process or 
bureaucratic measures, which may result in ineffective 
practices~\cite{Sutton2000}. The limitation of resources leads to a focus on 
product development instead of establishing rigid processes~\cite{Coleman2008, 
Heitlager2007}. Attempts to tailor lightweight processes to startups reported 
failures: ``Everyone is busy, and software engineering practices are often
one of the first places developers cut corners''~\cite{Martin2007}. Rejecting
the notion of repeatable and controlled processes, startups prominently take
advantage of reactive and low-precision~\cite{surviving-os-cockburn} engineering 
practices~\cite{Sutton2000,Tanabian2005,Chorev2006,Kakati2003}.

Startups typically develop software services that are licensed to customers 
rather than products that are sold and customized to a particular 
client~\cite{genome2012}. Market-driven software development (sometimes called 
packaged software development or COTS software development~\cite{regnell2001}) 
addresses issues related to this aspect.
Researchers emphasize the importance of time-to-market as a key
strategic objective~\cite{dagMDR,sawyer99} for companies operating in this
sector. Furthermore, requirements are ``invented by the software 
company''~\cite{512553}, ``rarely 
documented''~\cite{Karlsson02challengesin}, and can be validated only after the 
product is released to market~\cite{dahl2003,Keil:1995}. Hence, failed product 
launches are largely due to ``products not meeting customer 
needs''~\cite{Alves2006}.
To address this issue, startups embrace product-oriented practices with 
flexible teams, applying workflows that provide the ability to quickly 
change direction to the targeted market~\cite{Heitlager2007,Sutton2000}. 
Therefore, many startups focus on team productivity, granting more freedom to 
the employees instead of providing them with rigid 
guidelines~\cite{Tanabian2005,Chorev2006, Kakati2003}.

Can the goals of startups, namely accelerating time-to-market and meeting 
customer needs, be improved by the use of solid engineering practices 
customized for startups? Even though this specific question is not the focus of 
the study presented in this paper, the detailed investigation of 
state-of-practice is a prerequisite for future research into enabling the 
engineering taking place in startups.

\subsection{General lack of research in startups} 
Sutton~\cite{Sutton2000} noted in 2000 a general lack of studies in 
this area, claiming that ``software startups represent a segment that has been 
mostly neglected in process studies''. Further evidence for this observation 
is provided by Coleman and O'Connor~\cite{Coleman2008,Coleman2008a,Coleman2007} 
in 2008.
A Systematic Mapping Study (SMS)~\cite{SMS} performed in 2013 identified only a 
few studies into software engineering practices with focus on startups. 
Moreover, the identified studies are highly fragmented and spread across 
different areas rather than constituting a consistent body of knowledge. The 
following subsections discuss the findings of the SMS.

\subsection{Software development in startups}

Carmel~\cite{Camel1994a} introduced the term \textit{startup} to the SE 
literature in 1994, studying the time-to-completion in a young package firm. He 
noticed how these companies were particularly innovative and successful, 
advocating research to investigate their software development practices 
and enabling replication of their success by transferring their practices to 
other technology sectors.

Software startups are product-oriented in the first period of their development 
phase~\cite{Heitlager2007}. Despite good early achievements, software 
development and organizational management increase in 
complexity~\cite{1456074,Banker1998} causing deterioration of performance over 
time. Briefly, the necessity of establishing initial repeatable and scalable 
processes cannot be postponed forever~\cite{Ambler2002}. Starting
without any established workflows~\cite{Kajko-Mattsson2008}, startups grow over 
time, creating and stabilizing processes to eventually improve them only when 
sufficiently mature~\cite{Crowne2002}.

As startups have little time for training activities, as
discussed by Sutton~\cite{Sutton2000}, the focus shifts from prescriptive 
processes to team capabilities, hiring people who can ``hit the
ground running''~\cite{Yoffie1999}. Empowering the team and focusing on
methodological attributes of the processes oriented towards prototyping, 
proof-of-concepts, mock-ups and demos, testing basic functionalities, have been 
the priority in startups~\cite{Camel1994a}. With the startups' growth, 
coordinated quality control and long-term planning processes become 
necessary~\cite{Yoffie1999}.

Tingling~\cite{Tingling2007} studied the extent to which maturity of a 
company affects process adoption. He reports on introducing Extreme Programming 
(XP) principles~\cite{Beck:2004:EPE:1076267} in the development process, and 
the challenges arising from the need of trained team-members to fully 
implement the methodology. Similarly, da Silva and Kon~\cite{Silva2005} were
only able to start with all the XP practices in place after six months of
coaching the team. Nevertheless, even then, customization of practices need to 
be implemented, adapting the processes to the startups’ context~\cite{Deias}.

Contributions to flexibility and reactiveness of the development process exist 
by means of Lean~\cite{Gautam2008} and Agile~\cite{Abrahamsson2002} 
methodologies (also reported in~\cite{Taipale2010,Kuvinka2011}).
Startups face uncertain conditions, leading to a fast learning from trial 
and error, with a strong customer relationship, and avoiding wasting time in 
building unneeded functionality and preventing exhaustion of 
resources~\cite{Midler2008,Hilmola2003, Sutton2000}. Customer involvement in 
software development has also been discussed by Yogendra~\cite{Yogendra2002} as 
an important factor to encourage an early alignment of business concerns to 
technology strategies.

However, the question remains, to what extent can improved practices
in e.g. requirements engineering contribute to shortening time-to-market
or improve target market accuracy. There have been initiatives to
optimize practices for a specific purpose. McPhee and Eberlein~\cite{McPhee2002}
introduced practices adapted for reducing time-to-market.
Cohen et al. looked at development performance and time-to-market
trade-off~\cite{cohen1996}. None of these studies focus on startups per se,
but show that there is current knowledge that could be useful for
startups, or at least can function as a starting point for performing research
into solutions for startups.

In conclusion, since ``all decisions related to product development
are trade-off situations''~\cite{Hilmola2003}, startups generally optimize
workflows to the dynamic context they are involved in. Startups typically adopt 
any development style that might work to support their first needs, following 
the ``Just do it'' credo~\cite{Ries2011}. As remarked by Coleman and 
O'Connor~\cite{Coleman2008}, ``many managers just decide to apply what they 
know, as their experience tells them it is merely common sense''. This, 
however, does not preclude the possibility to collect, package and 
transfer experience in a lightweight manner, that allows flexible adoption of 
good engineering practices. On the contrary, startups that cannot benefit from 
very experienced team members would increase their success potential by 
following validated work practices.

\subsection{Software process improvement in startups}

The problem of one-size-fits-all, related to some SPI representations for
startups, is described by Fayad~\cite{Fayad1997}. He discusses the
problem in actuating the same best-practices criteria for established companies
in 10-person software startups. Sutton~\cite{Sutton2000} remarks that problems
of rigid SPI models in software startups arise due to: the dynamic nature of the development
process, which precludes repeatability; organizational maturity, which cannot be
maintained by startups lacking corporate direction; severe lack of
resources, both human and technological for process definition, implementation,
management, and training. In conclusion, the primary benefits of 
one-size-fits-all SPI often do not hold for startups, which instead of 
promoting product quality, aim to minimize time-to-market.

Additionally, the role of rigid SPI has been neglected because it is seen as an
obstacle to the team's creativity and flexibility, and to the need
of a quick product delivery process environment~\cite{Coleman2008a}. 
Product quality is often left aside in favor of minimal and suitable 
functionalities, shortening time-to-market. Mater and 
Subramanian~\cite{Mater2000} and Mirel~\cite{Mirel2000} report that the quality 
aspects mostly taken in consideration in internet startups are oriented towards 
usability and scalability. However, market and application type heavily
influence the quality demand~\cite{Coleman2008, Kim2005}.

To maintain the development activities, oriented towards limited but suitable
functionality, studies suggest externalizing the complexity of parts of the
project to third party solutions by outsourcing
activities~\cite{Hanna2010}, software reuse~\cite{Jansen2008} and open-source
strategies~\cite{Wall2001, Bean2005}.

\subsection{Technical debt} 
A new stream of SE research, trying to tackle the problem of technical 
debt~\cite{Tom2013}, brings and encompasses various implications in studying
development in software startups. The metaphoric neologism of technical debt was
originally introduced by Cunningham in 1992~\cite{TechnicalDebtCunn} and has
recently attracted the attention of SE researchers\footnote{Important
contributions characterizing the ``debt landscape'' 
are~\cite{Nugroho2011,Izurieta2012} published at a dedicated 
workshop~\cite{workshopDebt}
organized by the Software Engineering Institute and ICSE.}. Brown et 
al.~\cite{Brown:2010:MTD:1882362.1882373} provides an illustration of the 
technical debt concept:
``The idea is that developers sometimes accept compromises in a system in one
aspect (e.g., modularity) to meet an urgent demand in some other aspects
(e.g., a deadline), and that such compromises incur a ``debt'' on which
``interest'' has to be paid and which the ``principal'' should be repaid at some
point for the long-term health of the project''. Tom et al.~\cite{Tom2013} 
identified five dimensions of technical debt: code, design and architecture, 
environment, knowledge distribution and documentation, and testing.
On a daily basis startups face a trade-off between high-speed 
and high-quality engineering, not only in architecture design but in 
multifaceted aspects (weak project management, 
testing, process control). In the context of early-stage startups, we 
illustrate empirical evidence on accumulated technical debt in
subsection~\ref{res:gsm:cat5} and discuss its 
implications in subsection~\ref{sect:theory:impl:paydebt}.

\subsection{Terminology} To set a common ground and to prevent
ambiguity, we use the following terminology throughout the paper:

\begin{compactitem} 
\item Software development strategy: the overall approach adopted by the 
company to carry out product development.
\item Engineering activities: the activities needed to bring a product from 
idea to market. Traditional engineering activities are, among others,
requirement engineering, design, architecture, implementation, testing.
\item Engineering elements: any practice, tool or artifacts contributing to and 
supporting the engineering activities.
\item Quality attributes: those overall factors that affect run-time
behavior, system design, and user experience. They represent areas of concern
that have the potential for applications to impact across various layers and 
tiers. Some of these attributes are related to the overall system design, while 
others are specific to run time, design time, or user centric 
issues~\cite{Microsoft2009}.
\item Growth: an increase in company size with respect to the initial 
conditions for either employees or users/customers, and product
complexity for handling an increasing number of feature requests.
\item Software product: any software product and/or software service.
\item Software process improvement: any framework, practice, or tool that 
supports activities leading to a better software development 
process~\cite{5728832}.
\end{compactitem}

\section{Research methodology} \label{resmet}

\begin{figure*}[!t] \centering
\includegraphics[width=6in]{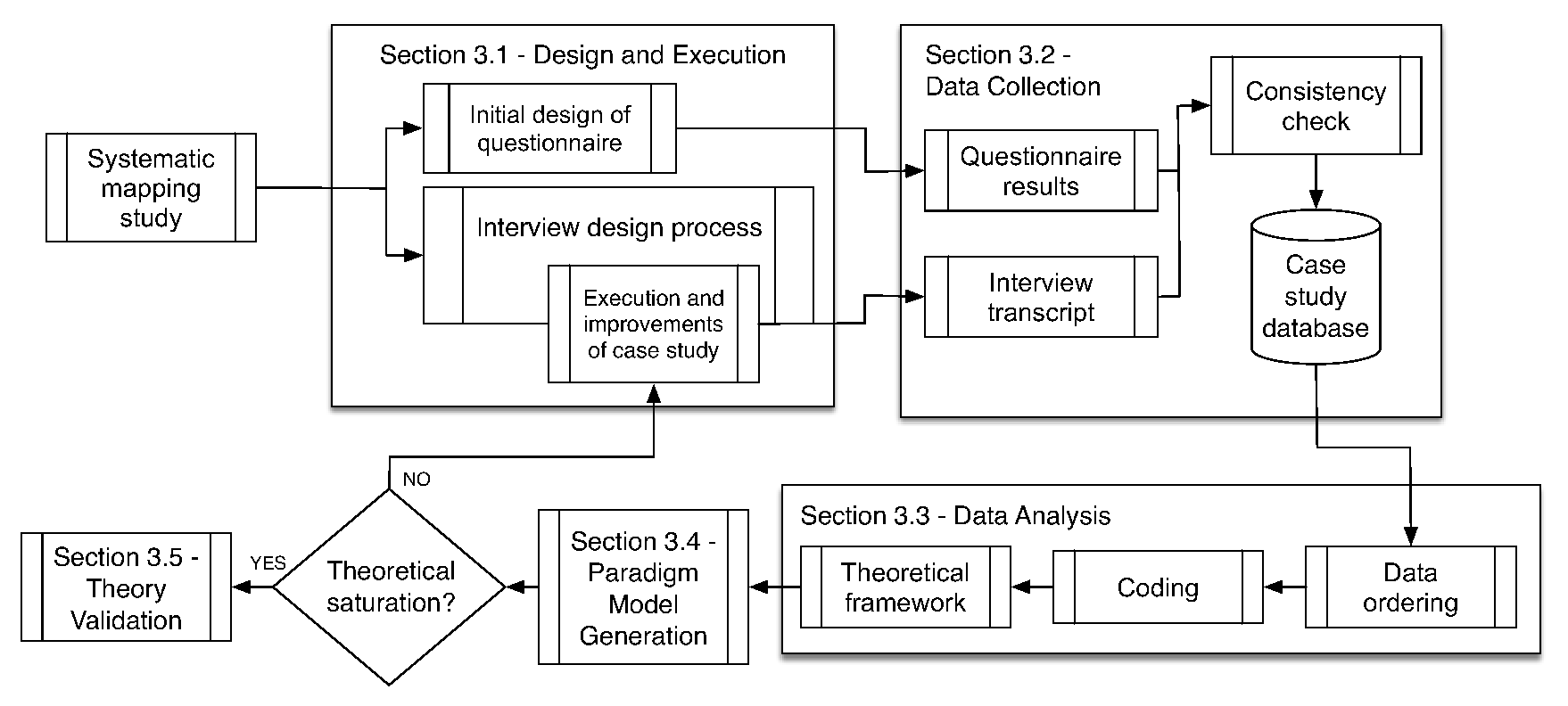} 
\caption{Research methodology - Grounded Theory process overview} 
\label{fig:gt:completemethodology} \end{figure*}

The goal of this study is to understand how software development strategies
are engineered by practitioners in startup companies. In particular, we are 
interested in structure, planning and control of software projects, in the 
period from idea conception to the first open beta release of the software 
product.

We set the boundaries of the research by reusing a previously conducted 
systematic mapping study~\cite{SMS}, which steered also the formulation of 
research questions:
\begin{compactitem}
  \item[RQ-1:] How do startups structure and execute their main engineering
    activities?
  \item[RQ-2:] How are product quality attributes considered by startups?
\end{compactitem}

To answer these questions, we investigated the software development approach 
undertaken by practitioners of startups. Following a Grounded Theory (GT) 
methodology~\cite{Glaser1978}, we
executed 13 semi-structured interviews (with 13 companies) integrated with
follow-up questionnaires. We tailored the questionnaires to each startup,
partially taking advantage of the repertory grid principles~\cite{Edwards2009}.
From this, we elaborated and extracted the Greenfield Startup Model (GSM)
explaining the underlying phenomenon of software development in startups.

Following the GT principles, we captured the most relevant aspects of software
development from startup practitioners, letting a theory emerge from the
interviews and adjusting the research hypotheses and questions as we proceeded.
During these interviews we collected data related to engineering activities
undertaken by startups. Then, we proceeded with the analysis of the data, 
finding important relations among concepts with a formal approach to
generate and validate the final theory~\cite{Glaser1978}.

As suggested by Coleman, in view of the different versions of GT, researchers
should indicate which ``implementation'' of the theory is being
used~\cite{Coleman2007}. Since information obtained from the SMS 
and our direct experience with startup companies provided a
good initial level of knowledge, in this study we use Corbin and Strauss'
approach~\cite{Strauss1998}. This GT version empowers the researchers' 
``theoretical sensitivity''~\cite{Corbin1990}, and encourages them to outline 
the research problem beforehand.

Figure~\ref{fig:gt:completemethodology} shows a complete overview of the study 
methodology and execution, illustrating how we tailored the general GT 
methodology to our specific needs. The produced data collection and analysis 
packages (including interview questions, follow-up questionnaires and codes) 
are available in the supplemental material of this 
paper~\cite{giardino_supplementary_2015}.

The results of our previous SMS provide input to the study design, contributing 
to the \textit{Design and Execution} of the study. The process depicted in
Figure~\ref{fig:gt:completemethodology} is evolutionary and affects the design
at each new iteration. In \textit{Data Collection} we integrate the empirical 
results in a case study database and subsequently process it in \textit{Data 
Analysis} to form theoretical categories. At each iteration, the emergent 
theory is updated following a formal procedure, \textit{Paradigm Model 
Generation}, and after verifying that we achieved \textit{Theoretical 
Saturation}\footnote{The point at which executing more interviews would not 
bring any additional value for constructing the theory.} of categories, we 
proceeded to \textit{Theory Validation}.

The first two authors jointly executed the whole procedure, handling
conflicts by reviewing the rationale of decisions with the third and fourth
authors. When necessary we performed an in-depth review of the study design  
and data collected during the execution process. The process details are 
described in the following subsections, structured according to the five macro 
phases depicted in Figure~\ref{fig:gt:completemethodology}.

\subsection{Design and Execution} \label{desex}
In this paper we address technical aspects related to software development in
startups, exploring their operational dynamics. Lacking agreement on a unique 
definition of the term \textit{startup}, we sampled case companies according to 
the recurrent themes characterized in the definition of startups~\cite{SMS}: 
\begin{compactitem}
 \item newly created: with little or no operating history.
 \item lack of resources: with economical, human, and physical limited resources.
 \item uncertainty: with little knowledge of the ecosystem under different 
perspectives: market, product features, competition, people and finance.
\item aiming to grow: with a scalable business in increasing number of users, 
customers and company's size.
\end{compactitem}

We sampled the companies in two distinct phases. First we executed an initial 
convenience sampling~\cite{Dawson2009}, which led to the identification of 
eight companies. Then we included five additional startups during the theory 
formation process (theoretical sampling), iteratively improving the sample 
according to the emerging theory. The characteristics of the sampled companies 
are reported in Table~\ref{t_interviews-stats}.

\begin{table}
\caption{Characteristics of the studied companies}
\label{t_interviews-stats}
\centering
\footnotesize
\begin{tabular}{cccccc}
\midrule
\midrule
    ID & Company age & Founding & Current & First product \\ & in months & team
    & employees & building time \\ & & (developers) & & in months \\
   \midrule
C1 & 11 & 4 (2) & 11 & 6 \\ C2 & 5 & 2 (2) & 6 & 3 \\ C3 & 18 & 4 (4) & 4 & 6 \\
C4 & 17 & 3 (2) & 11 & 6 \\ C5 & 20 & 2 (1) & 4 & 12 \\ C6 & 30 & 3 (2) & 4 & 1
\\ C7 & 12 & 2 (1) & 7 & 4 \\ C8 & 24 & 4 (3) & 16 & 4 \\ C9 & 5 & 5 (4) & 5 & 1
\\ C10 & 43 & 6 (4) & 9 & 4 \\ C11 & 36 & 3 (3) & 6 & 2 \\ C12 & 12 & 3 (3) & 3
& 3 \\ C13 & 24 & 2 (2) & 20 & 3 \\
 
\bottomrule
\end{tabular}
\end{table}

All companies, except C10, were founded within the last three years (2009-2012),
by an average of 3 founding members, who were in majority developers.
Moreover, the number of current employees shows how, to different degrees,
companies expanded the initial teams. All companies, except C5, released their
first product to the market within 6 months of the idea conception. The
products consist of pure web (8), web- and mobile (4), and web- and 
desktop applications (1), launched in six different nations (United States (4), 
Italy (4), Germany (2), Sweden (1), United Kingdom (1), New Zealand (1)). The 
growing team size and publicly available data suggest a generally healthy 
status of the businesses. A detailed documentation about the startup sampling 
and their distribution can be found in the supplemental material of this 
paper~\cite{giardino_supplementary_2015}. 
We executed the case studies online, supported by tools for video conferencing, 
recording each session which 
lasted 1 hour on average. The interview subjects were CEOs or CTOs. When 
selecting interviewees, we required that they worked at the company from the 
start. We followed a step-by-step work-flow, consisting of the actual 
interview, preparation of the customized follow-up questionnaire and the 
iterative adjustment of the interview package artifacts. 

\subsection{Data collection} 
We designed the data collection to allow for triangulation, which integrates
multiple data sources (interview, questionnaire) converging on the same 
phenomenon. The interview questions (see Table~10 in the supplemental  
material~\cite{giardino_supplementary_2015}) cover aspects such as development 
process, requirements elicitation, quality requirements, analysis, design, 
implementation, testing and deployment. After transcribing an interview, we 
sent a follow-up questionnaire to the interviewee. We designed the 
questionnaire to capture additional data, gather missing information and 
confirm interview results by triangulation. Note that we did not use the data 
from the follow-up questionnaire as input for theory generation. Table~11 in 
the supplemental material shows the prototype of the questionnaire that we 
adapted to each interviewee and company, based on the data collected in the 
earlier interview. 

The case study database allowed us to easily retrieve and
search for information, assembling the evidence from different data sources, as
described also by Yin~\cite{Yin1994}. We constructed and stored the database 
using the qualitative data analysis software package AtlasTI\footnote{Available
online at \url{http://www.atlasti.com/}.}. We overlapped interviews with 
questionnaire results to reveal and flag potential inconsistencies in the data.

\subsection{Data analysis}
\label{data_analysis} 
The first two authors led the coding procedure and performed the analysis in a 
co-located environment, i.e. working together on a single computer screen. 
Before starting the analysis, a data ordering procedure was necessary as
interviews were spread across a multitude of topics. 
Therefore, we structured the transcripts into thematic areas according to 
different topic cards used during the interviews. We proceeded horizontally to 
analyze the same thematic areas within different transcripts, rather than going 
through an entire transcript at one time. Once the data was ordered, we
coded the interviews according the following steps:

\begin{compactitem}
\item We assigned labels to raw data, and carried out a first low-level 
conceptualization using both in-vivo and open coding~\cite{ColinRobson2009}.
\item We grouped concepts together into theoretical categories and 
subcategories. By means of axial coding we first described the different
relations between subcategories, and then relations between subcategories and
categories.
\item We refined categories several times to create different levels of 
abstraction and adjusting concepts, aided by a simple knowledge management tool.
\item We validated consistency among categories by selective coding, exploring 
and analyzing links among subcategories.
\item We identified the core category - the one with the greatest explanatory 
power - by analyzing the causal relations between high-level categories.
\end{compactitem}

During data extraction we used in-vivo coding combined with the more 
descriptive procedure of open coding. 
Following the example of other grounded theories, developed in
related areas such as Information Systems~\cite{Orlikowski1993} and Software
Process Improvement~\cite{Coleman2006}, we performed the high-level
conceptualization during creation of categories, in the process of refining
axial and selective coding. As we were iterating through the interviews, we 
analyzed new data by updating codes and categories when 
necessary, and taking notes in the form of memos to adjust the emerging theory. 

After the coding process, we formalized a first representation of the GT 
experience map in a theoretical model. The model is presented in the form of 
categories and subcategories that are linked together according to cause-effect 
relationships~\cite{Corbin1990}. The formation of the theoretical model is a 
bottom-up approach. From the empirical data and coding process, the model 
developed into two different levels: a detailed level representing the network 
of subcategories (identified mainly by the axial coding process), and a 
high-level representation of the main categories network (identified mainly by 
the selective coding process).

\subsection{Paradigm model generation}

As mentioned in subsection~\ref{desex}, we tested emergent theories by
integrating additional companies into the sample, selected following the
principle of theoretical sampling~\cite{Yin1994}.

We used the process of paradigm modeling, introduced by 
Corbin~\cite{Corbin1990}, at each iteration together with interview execution, 
systematically analyzing the emerging theory. The paradigm model is composed of:

\begin{compactitem} 
\item Causal conditions: the events which lead to the occurrence of 
the phenomenon, that is our core category.
\item Context: set of conditions in which the phenomenon can be 
extrapolated.
\item Intervening conditions: the broader set of conditions with which
the phenomenon can be generalized.
\item Action/interaction strategies: the actions and responses that 
occur as the result of the phenomenon.
\item Consequences: specification of the outcomes, both intended and
unintended of the actions and interaction strategies.
\end{compactitem}

Within the limits of the critical bounding assumptions, the role of the
generated theory is to explain, predict and understand the underlying
phenomenon.

\subsection{Theory Validation} \label{rm:val} 

Presenting a grounded theory (GT) is challenging for a researcher, who must pay 
attention to structure the included level of detail, and to the way data is 
portrayed displaying evidence of emergent categories. To assess our study and 
to determine whether the GT is sufficiently grounded, we used a 
systematic technique to validate the theory. Strauss and Corbin provided a list 
of questions to assist in determining how well the findings are 
grounded~\cite{Strauss1998}:

\begin{compactenum} 
\item[Q1] Are concepts generated, and are the concepts systematically related? 
\item[Q2] Are there many conceptual linkages and are the categories well 
developed? 
\item[Q3] Is variation\footnote{Variation refers to the variety of contexts to 
which the theory can be applied.} built into the theory and are the conditions 
under which variation can be found built into the study and explained? 
\item[Q4] Are the conditions under which variation can be found built into the 
study and explained? 
\item[Q5] Has the process been taken into account, and does the theory stand 
the test of time? 
\item[Q6] Do the theoretical findings seem significant, and to what extent? 
\end{compactenum}

In the remainder of this section, we illustrate how we answered these six 
questions. We generated the concepts according to the described coding process 
(Q1) and systematically related them through the use of a network diagram (Q2). 
At each iteration of the grounded theory process, we considered and examined a 
concept within different conditions and dimensions, trying to incorporate data 
from a broader range of practitioners (Q3).
We constructed all the linkages and categories by the use of Atlas.TI and
compared them according to the data analysis process. Moreover, we connected 
extensive explanations, in form of in-vivo statements as reported by 
practitioners, to the developed concepts (Q4).

We designed the research process in multiple steps, explaining the purpose and 
implementation of each. Thus, the same process together with the supplemental 
material of this paper~\cite{giardino_supplementary_2015} enables other 
researchers to replicate our study within similar contexts (Q5). Moreover, we 
performed a comparison with the state-of-art to validate the theory and to 
strengthen its applicability within a wider time-frame (Q6). By this comparison 
we highlight the areas which have been neglected by existing studies, providing 
possible directions for future studies (see 
subsections~\ref{sect:theory:validation:others} 
and~\ref{sect:theory:validation:sms}). Furthermore, we studied the
confounding factors which could interfere with the application of the GSM
(see subsection~\ref{sect:an:comp:cat-vs-literature:conf}).

\section{Results: Greenfield Startup Model}\label{res:gsm}
\begin{figure*}[!t] 
\centering 
  \includegraphics[width=6in]{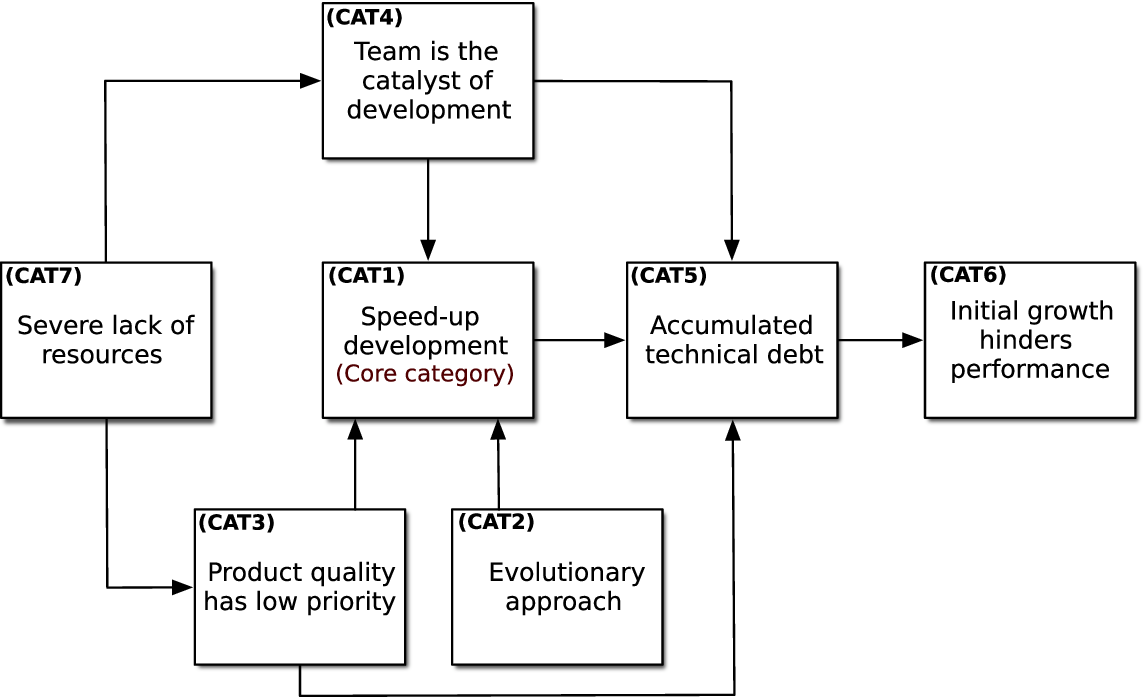}
  \caption{Main categories and causal relationships in the Greenfield Startup 
Model}\label{fig:gsm}
\end{figure*}

The GSM captures the underlying phenomenon of software development in 
early-stage startups. The model is formed by 128 sub-categories, clustered 
in 35 groups, and finally in 7 categories (see Figure~\ref{fig:gsm}) at the 
highest level of abstraction\footnote{All raw data, including codes, 
sub-categories and groups, are available in the supplemental 
material~\cite{giardino_supplementary_2015}.}. 
By the means of the GSM we provide explanations of the development strategies 
and engineering activities undertaken by startups. This section focuses on the 
data collected from the studied startups, forming the GSM. Note that in this 
section, we report on the GSM which is an abstraction of the collected 
empirical data from thirteen startups. The implications of the GSM and its 
validity are discussed in Sections~\ref{sect:theory:impl} and 
Section~\ref{res:val} respectively.

\subsection{Model overview} \label{res:gsm:frmov} 
We have grouped the main concepts representing the underlying phenomenon
together to form high-level categories. Figure~\ref{fig:gsm} shows the network 
of causal relationships (represented by arrows) between categories (represented 
by blocks).

In the forthcoming explanation of the GSM we make use of identifiers (i.e. CATx)
for the main categories shown in Figure~\ref{fig:gsm}. The network is centered
around the core category, \textit{speed up development}, which is the most
interconnected node in the theory reflecting the fact that ``it is the one
[category] with the greatest explanatory power''~\cite{Strauss1998}. 

A contextual condition, which characterizes to some extent every startup is 
the \textit{severe lack of resources}. In fact, limited access to human, time 
and intellectual resources constrain the capabilities of an early-stage startup 
to support its development activities. The \textit{severe lack of resources} 
forces the company to focus on implementing an essential set of 
functionalities. This is one of the main reasons why the \textit{product 
quality has low priority} with respect to other more urgent 
needs\footnote{There are some exceptions where the quality aspects
actually matter and such cases will be discussed in 
subsection~\ref{sect:an:comp:cat-vs-literature:conf}.}. At the same time, 
to be able to deal with such constraints, startups depend on a small group of 
capable and motivated individuals.

As unanimously expressed by respondents, the highest priority is to 
\textit{speed up the development} as much as possible by adopting a flexible 
and effective \textit{evolutionary approach}. The low attention given initially 
to architectural aspects related to product quality facilitates the efficiency 
of teamwork. This allows startups to have a functioning but faulty product, 
that can be quickly introduced to the market, starting from a prototype 
implementation on day-one.

The initial employees are the ingredients which enable high levels of
performance in software development. To support a fast-paced production
environment, engineers are required to be highly committed, co-located, 
multi-role, and self-organized. In other words, the \textit{team is the 
catalyst of development}. With an essential and flexible work-flow, which 
relies on tacit knowledge instead of formal documentation, startups can
achieve very short time-to-market cycles. However, each line of code, written 
without following structures and processes, contributes to growing the 
\textit{accumulated technical debt}, which is further increased by having 
almost non-existing specifications, a minimal project management and a lack of 
automated tests.

The consequences of such debt may not be perceived in the initial stages of a 
startup, where finding the product/market fit as quickly as possible is the 
most important priority. Startups, which survive to subsequent
phases will likely increase their user-base, product size, and number of 
developers. This will require the company to eventually pay the
\textit{accumulated technical debt}, and confront the fact that an
\textit{initial growth hinders productivity}.

In the following subsections we explain the categories presented in 
Figure~\ref{fig:gsm}, and conclude in subsection~\ref{res:gsm:th} with the
final theory. In the explanations we use identifiers of the companies presented
in Table~\ref{t_interviews-stats} (i.e. C1$...$C13) to highlight statements 
made by the interviewees.

\subsection{Severe lack of resources}\label{res:gsm:cat7} 
The concept of \textit{severe lack of resources} characterizes the uncertainty 
of development strategies in startups and it is composed of
three subcategories: \textit{time-shortage}, \textit{limited human resources} 
and \textit{limited access to expertise}.

Since startups want to bring the product to market as quickly as possible, the
resource they are the most deprived of is time. Startups operate under a
constant time pressure, mainly generated by external sources
(\textit{investor pressure}, \textit{business pressure}) and sometimes
internal necessities such as \textit{internal deadlines} and \textit{demo
presentations at events}. In this regard, C3 commented: ``Investors wanted to 
see product features, engineers wanted to make them better. Finally the 
time-to-market was considered more important and the teams' interests were
somehow sacrificed.''

In addition, to compensate for the \textit{limited human resources}, 
practitioners empower \textit{multi-role and full stack engineers}, as 
confirmed by C1: ``Everyone was involved in any tasks, from mobile to web 
development, organizing themselves in choosing the part to implement''. The 
extent to which startups have access to specialized knowledge - both
internal and external to the company - is reduced when compared to established 
software companies. Therefore, to partially mitigate the \textit{limited access 
to expertise}, startups rely on the external aid of mentors or advisors. Under 
these strict limitations, most of the decisions related to software development 
are fundamentally trade-off situations.

\subsection{Team as the development catalyst} \label{res:gsm:cat4} 
Among the different aspects fostering the speed of the development process, 
the startups' focus is on the characteristics of the initial team. In 
startups \textit{developers have big responsibilities}. In fact,
\textit{limited human resources}, discussed in CAT7, cause the team-members to
be active in every aspect of the development process, from the definition of
functionalities to the final deployment.

Engineers in the founding team of startups are sometimes \textit{multi-role} 
and typically \textit{full-stack engineers}. Multi-role engineers handle both 
the development and are at the same time responsible for marketing and sales. 
C1 observed that: ``A developer has many responsibilities, and needs to 
quickly move among a variety of tasks as there is no company hierarchy.''
Full-stack engineers can tackle different problems at various levels of 
the technology stack (\textit{generalist developers instead of specialists}).
C11 remarked that: ``Instead of hiring gurus in one technology, startups should 
hire young developers, generalists, that know how to quickly learn new 
technologies, and quickly move among a huge variety of tasks.''

Moreover, having a \textit{very small and co-located development team} enables
members to operate with high coordination, relying on \textit{tacit knowledge} 
and replacing most of the documentation with informal discussions. 
Practitioners reported that keeping the development
team small helps startups in being fast and flexible, as remarked by C8:
``If you have more than 10 people, it is absolutely impossible to be
fast''. Then, also \textit{basic knowledge of tools and standards of the 
working domain} and \textit{knowing each other before starting the company} 
support the efficiency of activities by \textit{limiting the need for 
formalities between team members}.

In every software company, \textit{skilled developers are essential for high
speed} development. Especially in startups, the ``hacking culture'' and a
tendency to the ``just-do-it'' approach allow the team to quickly move from the
formulation of a feature idea to its implementation. In this regard, C1
comments: ``We had a hacker culture/environment, people hacking stuff without 
formally analyzing it, but breaking it down and finding a way around.''

A \textit{limited access to expertise} forces the team to rely mainly on their
personal abilities, even though interviewees reported that asking mentors for 
an opinion is a viable practice to aim for feasible objectives. 
Furthermore, \textit{teams work under constant pressure} mainly constrained by 
a tight \textit{time shortage}.

Finally, startups present founders-centric structures, and especially in the
early-stage, the \textit{CTO/CEO background has high-impact} on the
company's development approach. For instance, in case of an academic background,
the CTO might encourage the introduction of some architectural design before the
development phase. Even though the CTO/CEO initially guides the development 
process, most of the decisions are taken collectively by all members of the 
team. Then, the CTO/CEO only intervenes in situations where conflicts occur.

\subsection{Evolutionary approach}\label{res:gsm:cat2} 
Startups prefer to build an initial prototype and iteratively refine it over 
time, similarly to the concept of ``evolutionary prototyping''~\cite{EvProt}. 
The goal is to validate the product in the market as soon as possible, finding 
the proper product/market fit. Indeed, startups can focus on developing only 
parts of the system they want to validate instead of working on developing a 
whole new system. Then, as the prototype is released, users detect 
opportunities for new functionalities and improvements, and provide their 
feedback to developers.

Since \textit{flexibility and reactiveness are the main priorities}, the most
suitable class of software development approaches are highly
evolutionary in nature. As \textit{uncertain conditions make long-term
planning not viable}, startups cannot base their work on assumptions without
rapidly validating them by releasing the product to market. Uncertainty lies
first of all in the team composition. Since the teams are typically small
and project knowledge is generally undocumented, even a minor change in their
composition (e.g. a developer falls ill) can have a significant impact on the
overall product development. Furthermore, startups operate in a continuously
evolving environment of competitors and targeted market sectors. Then,
to get a competitive advantage in the market, startups typically make use of
cutting-edge solutions, characterized by an evolution that cannot be foreseen in
the long run. However, user feedback and requests play a special role in daily 
decisions as main drivers for defining the product features in the short term.

To obtain fast user responses and quickly \textit{validate the product},
startups \textit{build a functioning prototype and iterate it} over time.
Quoting C4, ``[\ldots] you should start with something that is
really rough and then polish it, fix it and iterate. We were under constant
pressure. The aim was to understand as soon as possible the product market/fit
iterating quickly, adjusting the product and releasing fast.'' The companies 
focus on building a small set of functionalities to include in
the first version, and \textit{progressively roll-out to a larger number of
people} with \textit{small iterations} (confirmed by C4: ``we deploy 
from 5 to 20 times a day'').

The objective of this evolutionary approach is to avoid wasting time on 
``over-engineering the system'' and building complex functionalities that have 
not been tested on real users. By releasing a small number of good-enough
functionalities (see CAT3) the startup verifies the suitability of the
features and understands how to adjust the direction of product development
towards actual users' needs. The first version of the product is typically a
prototype containing basic functionalities developed with the least
possible effort that validates critical features, enabling the startup's
survival in the short term. Supported by \textit{direct contact and 
observation of users}, \textit{automated feedback collection} and 
\textit{analysis of product metrics}, startups attempt to \textit{find what is 
valuable for customers}.

\subsection{Product quality has low priority}\label{res:gsm:cat3}
The interests of software startups, related to the product, are  
concentrated on building a \textit{limited number of suitable functionalities} 
rather than fulfilling non-functional requirements. This strategy 
allows them to quickly release simple products with less need for preliminary 
architectural studies.

The quality aspects considered by startups during the development process are 
geared towards user experience (UX\footnote{According
to ISO 9241-210 (Ergonomics of human-system interaction), UX is defined as
``a person's perceptions and responses that result from the use or
anticipated use of a product, system or service''.}), in particular 
\textit{ease of use, attractiveness of the UI} and \textit{smooth user-flow 
without interruptions}. C11 notes that \textit{UX is an important 
quality factor}: ``When a user needs to think too much on what action should 
be done next, he will just close the application without returning''. C3 adds: 
``If the product works, but it is not usable, it doesn't work''.

The extent to which quality aspects are taken into account might depend on 
the market sector and the type of application. Nevertheless, realizing a high
level of UX is often the most important attribute to consider for customer
discovery of evolutionary approaches in view of the \textit{limited human
resources} and \textit{time shortage}, presented in CAT7. C4 confirms:
``None of the quality aspects matter that much as the development speed does.''

To achieve a good level of UX while dealing with lack of human resources and
time shortages, startups analyze similar products, developed by larger 
companies that can afford more rigorous usability studies. Then, the users' 
feedback and product metrics begin to have a central role in determining the 
achieved UX level. Product metrics are typically web-based statistical 
hypothesis testing, such as A/B testing~\cite{AB}. Other than UX, some other 
factors can influence the quality concerns of development:

\begin{compactitem}
\item The \textit{efficiency emerges after using the product}, letting
engineers avoid wasting time in excessive improvements of not-validated
functionalities. 
\item The \textit{product should be reasonably ready-to-scale} to be able to 
accommodate a potential growth of the user-base. Startups \textit{externalize 
complexity to third party solutions}, such as modern cloud services, achieving 
a sufficient level of scalability.
\item Realizing high reliability is not an urgent priority as \textit{users 
are fault-tolerant towards innovative beta products}. In these cases, users 
typically have a positive attitude towards the product, even though it exhibits 
unreliable behavior. In this regard, the focus of beta testing is 
reducing friction between the product and the users, often incorporating 
usability testing. In fact, the beta release is typically the first time that 
the software is available outside of the developing organization\footnote{A 
discussion of the impact of innovative products on the user 
satisfaction is presented in 
subsection~\ref{sect:an:comp:cat-vs-literature:conf}.}. 
\end{compactitem}

\subsection{Speed-up development}\label{res:gsm:cat1} 
\textit{Speed up development} represents the core 
category of the GSM. Firmly grounded as the primary objective of startups, it 
shows the most important characteristic of developing software in the early 
stages.

To \textit{speed up development}, startups adopt evolutionary 
approaches supported by a solid team focusing on implementing a
minimal set of suitable functionalities. Startups \textit{keep simple and
informal workflows} to be flexible and reactive, adapting to a fast changing
environment. The fact that teams are typically self-organized and 
\textit{developers have significant responsibilities} facilitates the adoption 
of informal workflows. The aim to shorten time-to-market restricts potential 
planning activities, as reported by C8: ``Speed was of essence so we didn't 
plan out too many details''. To deal with such unpredictability, startups 
prefer to take decisions as fast as possible, mainly by means of informal and
frequent verbal discussions.

Even though Agile principles embrace change, startups often perceive 
development practices as a waste of time and ignore them to accommodate the 
need for releasing the product to the market quickly. This 
approach is possible also in view of a lack of systematic quality assurance 
activities; startups focus on user experience and other quality aspects, such 
as efficiency, can be postponed until after the first release.

Another beneficial strategy that startups employ to quickly deliver products is 
the \textit{externalization of complexity on third party solutions}. Startups 
make use of third party components (COTS) and open source solutions (for 
product components, development tools and libraries). They take advantage of
external services for the sake of delivering a \textit{product reasonably ready 
to scale} for possible future growth. Moreover, advanced version control 
systems are not only used to manage the code-base, but also in task assignment, 
responsibility tracing, configuration and issue  management, automatic 
deployment, and informal code walkthroughs when issues occur. Even though 
\textit{the use of well-integrated and simple tools} allows startups to 
automate many activities and reduce their completion time, drawbacks of such 
approaches are increased interoperability issues. 

Startups further improve development speed by making \textit{use of
standards and known technologies} which are widely recognized, well tested, and
supported by strong communities. Moreover, the use of standards and frameworks 
reduces the need for a formal architectural design since most of the solutions 
are well documented and ready-to-use. C1 stated that: ``as long as you use Ruby 
standards with the Rails framework, the language is clean itself and doesn't 
need much documentation''.

Other important factors that positively impact the speed of development are the
team's desire to \textit{create disruptive technologies}, to
\textit{demonstrate personal abilities}, and to \textit{have the product used in
the market}. As reported by practitioners, these factors are essential to
enhance the morale of developers and therefore to achieve higher team
performance. On the other hand, when a developer is not \textit{able to meet 
deadlines}, especially in the typical sprint-based environments of Agile, the 
morale goes down, hindering the development speed.

Finally, the constant pressure under which the company regularly operates, leads
the team to often \textit{work overtime to meet deadlines}. But as reported by
practitioners, such a way of working can be an effective strategy only in the
short term since it can lead to poorly maintainable code and developer burnout
in the long run.

\subsection{Accumulated technical debt}\label{res:gsm:cat5} 
Startups achieve high development speed by radically ignoring aspects 
related to documentation, structures and processes. C4 stated that: ``You have 
to accept some extent of technical debt and some flawed code so you can move 
faster. You have to hit the sweet spot of moving very fast but at the same time 
without writing code that is so bad that you can’t update it anymore.''

Instead of traditional requirement engineering activities, startups make use of
\textit{informal specification of functionalities} through ticket-based
tools to manage low-precision lists of features to implement, written in the 
form of self-explanatory user stories~\cite{AgilePlan}. Practitioners 
intensively use physical tools such as post-it notes and whiteboards, which 
help in making functionalities visible and prioritizing stories based on 
personal experiences. C4 commented that ``[\ldots] it is the only way. Too many 
people make the mistake of sitting down and write big specs and then they build 
it for four months, realizing the product is not valuable only at the end.''

Since startups are risky businesses by nature, even less attention is given to
the traditional phase of analysis, which they replace by a \textit{rough
and quick feasibility study}. However, this approach has also disadvantages, as 
observed by C7: ``Some months later I started realizing the drawbacks: now that 
we have to grow, it would be nice to have done some more detailed study\ldots 
But at the same time, maybe if I did the study, I wouldn't  have all the 
agility and flexibility that we have now. It’s a big tradeoff.'' It is 
generally \textit{hard to analyze risks with cutting-edge technologies}. 
To find out the feasibility of such cutting-edge projects, startups attempt a 
first implementation with rough and informal specifications, assuming that the 
project's complexity will remain limited to a few functionalities, as discussed 
in CAT3 (subsection~\ref{res:gsm:cat3}). Additionally, by \textit{keeping the 
product as simple as possible} and learning from competitors' solutions and 
mistakes, practitioners use their \textit{past experiences in similar contexts 
to help to assess the feasibility} of the project.
Finally, to avoid restrictions on the flexibility of the team, potentially
limiting decisions are taken only when strictly necessary and as late as 
possible. Limiting, early decisions can increase the technical debt as 
commented by C8: ``Our biggest shortcoming was a poor initial decision on data 
structuring which was fundamental as the whole code (and the business logic) 
relied on it.  95\% was right, and 5\% of the data structure was wrong, and 
caused a lot of troubles (refactoring and re-doing code).''

Another important factor that contributes to \textit{the accumulation of 
technical debt} is the general \textit{lack of architectural design}, 
substituted by \textit{high-level mock-ups and low-precision diagrams},  
\textit{describing critical interactions with third-party components only}. In 
particular, the use of well-known standards, frameworks and conventions removes 
the need for formal UML~\cite{UML} diagrams and documentation, and provides a 
minimum level of maintenance costs. C6 stated that: ``\ldots with perfect 
hindsight we should have used a framework to create more maintainability of the 
code.  At the beginning, we didn't use the framework to develop the application 
faster. We believe that the additional time needed to use the framework would 
have payed off, because it would have increased understandability of the code 
structure and decrease the time needed for new developers to start working.'' 

A similar attitude towards verification and validation brings startups to a
\textit{lack of automated testing}, which is often replaced by manual
smoke tests. Quoting C3, ``Trying the product internally allows us to get 
rid of 50\% of bugs of important functionalities. Meanwhile, users report 
bugs of secondary functionalities, eventually allowing us to mitigate the lack 
of testing. Indeed, staying one week in production enables us to identify 90\% 
of bugs''. However, in certain cases where components of the system might cause 
loss of data or severe damages to the product or users, engineers realize a 
reasonable level of automatic testing. In such cases, aided by modern
automatic tools, they quickly assess the status of the system integration as
they add new functionalities to the product.

Startups perceive rigid project management as a ``waste of time'' that hinders 
development speed since the \textit{uncertainty makes
formal scheduling pointless} (C9 reported that ``initial chaos helps to
develop faster''). Startups' \textit{minimal project management} is supported
by keeping: \textit{internal milestones short and informal}, 
\textit{low-precision task assignment mechanisms} and a low cost 
project metrics (quoting C13, ``the only track of progress was made by 
looking at closed tickets''). In this context \textit{only a final release 
milestone is viable}, which helps practitioners to remain focused on short term
goals and put new features in production. 

Finally, one of the categories that contributes most to growing
\textit{accumulated technical debt} is the substantial use of informal and
verbal communication channels on a daily basis. The high co-location and the
fast paced development approach increase the volume of \textit{tacit knowledge}
and the severe lack of any kind of documentation. C4 observed in this regard 
that: ``[\ldots] the issue of having documentation and diagrams out of the
source code is that you need to update them every time you change something.
There is no time for it. Instead, there is a huge pay off in having a code that
is understandable itself.'' On the other hand, there are situations where this 
strategy is not good enough, as observed by C1: ``I had problems due to the 
lack of documentation. The only back-end documentation was the front 
end-design, so I had to guess what was behind!''.

\subsection{Initial growth hinders performance}\label{res:gsm:cat6} 
The lack of attention given in the first phases to engineering activities 
allows startups to ship code quickly. 
However, if the startup survives, the initial product becomes more complex over 
time, the number of users increases and the company starts to grow. Under these 
circumstances the need to control the initial chaos forces the development 
team to return the \textit{accumulated technical debt}, instead of focusing on 
new users' requests. Hence, the \textit{initial growth hinders performance} in 
terms of new functionalities delivered to the users.

When the user base increases, customers become more quality demanding and
scalability issues might start to arise. \textit{Company and user size
grow} when business events occur, such as: a \textit{new round of funding},
a possible \textit{acquisition}, the release of a \textit{competing product on
the market}, or when the project is \textit{open for the first public release}.
Therefore, while the project lacks even minimal processes, \textit{the current 
team is not able to manage increased complexity} of new functionalities and 
maintain the codebase.

Subsequently, practitioners start considering the need for project management
activities, also in view of \textit{hiring new staff members}, as discussed by
C13: ``[Project management] is strictly necessary if you radically
change the team or when the team grows. The informal communication and lack of
documentation slow down the process afterwards''. Project management becomes
even more important when the \textit{focus shifts to business concerns}. Part of
the effort, which was initially almost entirely dedicated to product
development, moves to business activities. Moreover, the availability of 
project information becomes an important issue as the accumulated \textit{tacit 
knowledge} hinders the ability of new hires to start working on project tasks.

Another factor that slows down performance is that \textit{portions of code 
need to be rewritten} and \textit{substantial refactoring of the codebase} is 
required by increasing product demands.
Practitioners realized that some decisions taken (or not taken) during the
\textit{rough and quick feasibility study} before starting the implementation,
have led to negative consequences on the long term performance and
maintainability of the product. The combination of these factors leads to the 
need to \textit{re-engineer the product}.
By re-engineering the systems, startups aim to \textit{increase the scalability
of the product/infrastructure} and start to \textit{standardize the codebase
with well-known frameworks}. C7 reports that: ``To mitigate this (lack of 
frameworks) I had to make a schema for other developers when we hired them. We 
had to do a big refactoring of the codebase, moving it from custom php to 
Django, normalizing the model and making it stick with the business strategy. I 
had the code in different php servers communicating via JSON, some engineering 
horror. Now that we are fixing it, it's really painful. We had to trash some 
code. However I don't regret that I didn't make this choice sooner, it was the 
only way''.

The \textit{fear of changing a product, which is working}, arises when product
complexity increases. The changes to the codebase, to support bug fixing, become
highly interrelated with other functionalities and difficult to manage because
the product is poorly engineered. Therefore, the fear arises that
changing a validated product might cause changes to users' responses.
The increasing number of feature requests leads to the \textit{growing necessity
of having a release plan}. Therefore, startups begin to \textit{partially 
replace informal communication with traceable systems} and \textit{introduce 
basic metrics for measuring project and team progress} to establish an initial 
structured workflow. Yet, C11 stated that: ``[\ldots] it is still better to 
have a reasonable drop-down in performance when the team grows than lose time 
in the beginning''.

\subsection{Paradigm model}\label{res:gsm:th}
To explain and understand the development strategies in early-stage 
software startups we construct the theory generated and supported by the above 
presented GSM:

\begin{theory} Focusing on a limited number of suitable functionalities, and
adopting partial and rapid evolutionary development approaches, early-stage
software startups operate at high development speed, aided by skilled and
highly co-located developers. Through these development strategies, early-stage
software startups aim to find early product/market fit within uncertain 
conditions and severe lack of resources. However, by speeding-up
the development process, they accumulate technical debt, causing an initial and
temporary drop-down in performance before setting off for further growth.
\end{theory}

We formed this theory by considering the different elements specified by 
Corbin~\cite{Corbin1990}:

\begin{compactitem}
\item ``Causal conditions'' are represented by three main conceptual
categories: \textit{product quality has low priority}, \textit{evolutionary
approach} and \textit{team is the catalyst of development}. 
\item ``Phenomenon'' is represented by the core category \textit{speed up
development}. 
\item ``Context'' is limited to early-stage web software startups 
operating in conditions of severe lack of resources aiming to early
find product/market fit. 
\item ``Intervening conditions'' are summarized by the extremely 
uncertain development environment. 
\item ``Action and interaction strategies'' are represented by the 
accumulation of technical debt.
\item ``Consequences'' lead to a temporary performance drop-off.
\end{compactitem}

\section{Implications of the GSM}
\label{sect:theory:impl}

In this section we present relevant implications that emerge from the behavior
of early-stage startups, formally expressed in the GSM. Although the startups 
we studied were spread across various nations and market sectors (see 
subsection~\ref{desex}), certain patterns emerged. We discuss these
patterns with respect to literature and identify possible venues for future 
research.

\subsection{Light-weight methodology}  
The most urgent priority of software development in startups is to shorten
time-to-market to find the right product/market fit. However, focusing on 
building and releasing the first version of a product, startups tend to not 
apply any specific or standard development methodologies or processes. Three 
interviewees (C5, C7, C13) referenced the Lean startup
methodology~\cite{Ries2011}, a highly evolutionary development approach,
centered around the quick production of a functioning prototype and guided by
customer feedback. However, none of the studied startups strictly followed the 
complete ``build-measure-learn'' cycle proposed by the Lean startup methodology.
One of the main purposes of Lean is waste reduction, although the 
identification of waste is not an easy matter as it spans perspectives and 
time~\cite{Poppendieck2006}. For example, running a value stream mapping is 
resource intensive, something that may put off startups. Nevertheless, even 
though the absence of a basic process might enable startups to focus
more on the product, startup companies can take advantage of some engineering 
activities even in the early stages~\cite{989006}. For instance, 
Taipale~\cite{Taipale2010} reports how startups benefited from tailoring some 
simple XP practices to their needs.

Startups in the early stage apply fast cycles of ``build and fix'' when 
necessary to act quickly and decisively enough to get the first response from 
the market. However, the  lack of perceivable cause and effect 
relationships constrains effective analysis~\cite{Kurtz2003}. Hence, applying 
best practices in a highly uncertain environment might be counter-productive. 
There is little to analyze yet, and waiting for patterns to emerge can be 
considered a waste of time. Quickly developing a set of suitable 
functionalities allows the team-members to present a prototype to a small set 
of potential customers and investors to start collecting quick feedback and 
respond accordingly. However, the studied startups do not 
explicitly follow the step-by-step process of ``customer development''  defined 
by Blank~\cite{Blank2005}. Instead, they absorb and implement the high-level 
principles from the customer development methodology, reflected in the GSM by 
the theoretical category \textit{find the product/market fit quickly}.

From a research perspective, collaboration with startups and technology 
transfer to those companies is challenging. State-of-the-art technology 
transfer models require long-term commitment from all 
participants~\cite{4012630}, an investment that might not be acceptable for 
an early-stage startup. Thus, there is a need to develop and validate 
technology transfer models adapted to the startup context.

\subsection{Empowering the team members}  
\label{sect:theory:impl:empteam}
The Lean startup methodology proposed by Ries~\cite{Ries2011} emphasizes 
team empowerment as a critical factor to pursue the development of a Minimum 
Viable Product (MVP). Empowerment allows the team to move rapidly and cut 
through the bureaucracy, approval committees and veto cultures. However, 
empowerment cannot be implemented without structure and means to measure 
performance~\cite{Randolph199519}. Startups can use lightweight tools, for 
example collection and evaluation of key performance indicators, task 
management and continuous deployment, to enable information sharing and 
autonomy creation which are key aspects of empowerment~\cite{Randolph199519}.

Yang~\cite{Team}, unlike to Ries' methodology, structurally differentiates four
dimensions that positively impact performance and should be considered in 
empowerment programs:
\begin{itemize}  
\item autonomy of taking decisions, where team-members can choose the  
activities they are interested in; 
\item responsibility for organizational results or success, keeping track of  
their own performance;  
\item information such that team members have influence on making decisions;
\item creativity, enabled by a culture where negative results are not 
punished, but attempts are rewarded;
\end{itemize}

Different forms of coordination methods utilize the idea of dividing problem 
and solutions space, like handshaking presented by Fricker et 
al.~\cite{Fricker2010}. These could also be investigated, especially since 
the main manager of a startup (CTO/CEO) cannot be involved in all solution 
decisions~\cite{Fricker2008}. 
Even though the GSM identifies and explains the startups' focus on 
characteristics of the initial team, further research is needed to adapt and 
validate team empowerment programs in the startup context that can foster the 
speed of development processes.

\subsection{Focus on minimal set of functionalities} 

To deliver a product with the right features built in, startups need to 
prioritize and filter. From an engineering point of view, most startups do not 
explicitly apply traditional Requirement Engineering (RE) activities to collect 
and manage requirements. However, by integrating simple techniques such as 
Persona and Scenario, companies can improve the effectiveness of requirements 
elicitation even with mostly unknown final users~\cite{1531030}, thereby also
shortening time-to-market. 

Another study suggests that using a lightweight
project-initiation framework such as the Agile Inception Deck can help in
preventing premature failure of the software project due to a wrong 
understanding of the project requirements~\cite{1667597}. 
Looking at RE in general, there are several good practice guidelines that 
are adapted for small organizations, where the organization can choose what is 
relevant for them, see e.g. uniREPM~\cite{Svahnberg2013}. The key is that even 
startups can benefit from a limited and fast inventory of good engineering 
practices.

\subsection{Paying back the technical debt}
\label{sect:theory:impl:paydebt}

To be faster, startups may use technical debt as an investment, whose 
repayment may never come due. Tom et al.~\cite{Tom2013} refer to ``debt 
amnesty'' as a written off debt when a feature or product fails.

Even though potentially useful in the short-term, over time technical debt has 
a negative impact on morale, productivity and product quality. Kruchten et 
al.~\cite{6336722} suggest identifying debt and its causes, e.g. by listing 
debt-related tasks in a common backlog during 
release and iteration planning. Tracking technical debt can also be conducted 
by measuring usability and scalability of the product, paying attention to the 
customers' behaviors through real-time and predictive 
monitoring~\cite{Ries2011}.

An alternative to control technical debt with small effort, as stated by many interviewees, is the use of modern coding platforms (e.g.
Github) and well-known frameworks. Coding platforms allow developers
to integrate several engineering activities such as requirements
lists, issue tracking, source control, documentation, continuous integration,
release and configuration management. Frameworks include support programs,
compilers, code libraries and tool sets to enable the initial development of a
project with limited overhead. However, these strategies target only particular 
dimensions~\cite{Tom2013} of technical debt, such as environmental and 
knowledge debt. 

Furthermore, to be effective in the selection of third party components and 
frameworks, startups need to perform an efficient impact analysis of
their process configuration. Technology selection frameworks have been used to 
stimulate innovation~\cite{Rohrbeck2006}, as decision making 
support~\cite{Shehabuddeen2006, Azzone2008}, and in tool 
selection~\cite{Aranda2006}. However, such approaches need to be adapted to the 
particular constraints and context of early-stage startups. 

\begin{table*}
	\caption{GSM categories mapped to concepts reported in related models}
	\label{tab:framework_comparison}
	\centering
	\footnotesize
	\begin{tabular}{p{1cm}p{16cm}}
		\toprule
		Category & Coleman~\cite{Coleman2007} \\ 
		\midrule
		\midrule
		CAT1 & Experience the lack of rigid engineering activities and 
		documentation. Flexibility and process erosion maintaining simple and 
		informal work-flows.\\
		CAT4 & CTOs' and CEOs' background has a great impact on the adopted 
		development process. Nevertheless, team members remain 
		self-organized, able to intervene in all the aspects of the development 
		process without any direct supervision.\\
		CAT5 & Verbal communication and lack of heavy documentation and 
		bureaucracy.\\
		CAT6 & Nimble and ad-hoc solutions prevent the use of heavy bureaucracy 
		and formal communication strategies, even though the accumulated tacit 
		knowledge is hard to manage and transfer to new hires.\\
		\midrule
		Category & Baskerville~\cite{Internet}\\ 
		\midrule
		\midrule
		CAT1 & Make heavy use of simple tools and existing components.\\
		CAT2 & Uncertain conditions make long-term planning not viable. 
		Speed-up development by releasing more often the software and 
		``implanting'' customers in the development environment.\\
		CAT3 & Tailor the development process daily according to the intense 
		demands for speed, skipping phases or tasks that might impede the 
		ability to deliver software quickly even though producing lower quality 
		software.\\
		CAT5 & Invest time in facilitating development of scalable systems by 
		the use of simple but stable architectural solutions.\\
		CAT7 & A desperate rush-to-market. A lack of experience developing 
		software under the conditions this environment imposes.\\
		\midrule
		Category & Brooks~\cite{BrooksJr1987}\\ 
		\midrule
		\midrule
		CAT1  &  The most radical possible solution for constructing software 
		is not to construct it at all, taking advantage of what others have 
		already implemented. It is the main strategy, which enables companies 
		to externalize complexity to third party solutions. \\
		CAT2 & Avoid deciding precisely  what to build but rather iteratively 
		extract and refine the product requirements  from customers and users. 
		\\
		CAT3 &  Starting from simple solutions allows  creating early 
		prototypes and control complexity over time. \\
		CAT4 & People are the center of a software project and it is  
		important to empower and liberate their creative mind.\\
		\bottomrule
	\end{tabular}
\end{table*}

\subsection{Synthesis} 
With slightly different levels of adherence, the presented implications are 
reflected in the behavior of most of the companies we studied. The 
results of this analysis indicate that early-stage startups are far from 
adopting standard development methodologies. The typical tendency is to
focus on the teams' capability to implement and quickly iterate on a prototype,
which is released very fast. Thus, in a context where it is hard for even the 
most lightweight agile methodologies to penetrate, research should focus on the 
trade-off between development speed and accumulated technical 
debt~\cite{Brown:2010:MTD:1882362.1882373}, which appears to be the most 
important determinant for the success of an early-stage startup.

Our investigation of early-stage startups opens up several opportunities for 
further research. Most importantly, the performance drop-down caused by the 
necessity of returning the accumulated technical debt while expanding the 
company's operations and structuring mitigation strategies needs to be 
addressed. This can be achieved by meeting the following four software 
development objectives:
\begin{itemize}
 \item integrating scalable solutions with fast iterations and a
minimal set of functionalities (this allows startups to maintain effective
planning and realistic expectations)
 \item empowering team members enabling them to operate horizontally in all the 
activities of the development environment simultaneously
 \item improve desirable workflow patterns through the initiation of a
minimal project management over time, as a natural result of emerging activities
of tracing project progress and task assignment mechanisms
 \item then, only when the chaos has been initially managed, planning 
long-term performance by adoption of Agile and Lean development practices.
\end{itemize} 
Eventually, to enable the introduction and adoption of new development 
methodologies, research is needed on new/adapted technology transfer models 
from academia and industry to startups' contexts.

\section{Theory Validation} \label{res:val}

In this section we discuss the validity of the GSM by means of 
cross-methodological observations, as discussed in subsection~\ref{rm:val}.
As we refer to the GSM's main categories throughout the validation, we list 
their name and corresponding subsection where they have been introduced:
\begin{compactitem}
\item[CAT1] Speed-up development (\ref{res:gsm:cat1}).
\item[CAT2] Evolutionary approach (\ref{res:gsm:cat2}).
\item[CAT3] Product quality has low priority (\ref{res:gsm:cat3}).
\item[CAT4] Team is catalyst of development speed (\ref{res:gsm:cat4}).
\item[CAT5] Accumulated technical debt (\ref{res:gsm:cat5}).
\item[CAT6] Initial growth hinders performance (\ref{res:gsm:cat6}).
\item[CAT7] Severe lack of resources (\ref{res:gsm:cat7}).
\end{compactitem}

\subsection{Comparison with other models}
\label{sect:theory:validation:others}

To validate the generalization of the model, we describe conceptualizations 
derived from the GSM that are supported by previous models developed by 
Coleman~\cite{Coleman2007,Coleman2008a, Coleman2008},  
Baskerville~\cite{Internet} and Brooks~\cite{BrooksJr1987}. 
Table~\ref{tab:framework_comparison} presents an overview of the 
comparison, mapping GSM categories to aspects reported in literature.

We refer to Coleman's work since he has conducted similar studies in 
the context of startups, even though with a different focus. Coleman 
investigated factors in software development that hinder initiatives of 
one-size-fits-all software process improvement (SPI) in a later stage, 
representing also companies in the expansion phase with more than 100
employees.

Coleman aims to highlight how managers consider two distinct kinds
of processes: \textit{essentials} and \textit{non-essentials}. The
essential processes are the most closely linked to product development,
such as requirements gathering, design and testing. The non-essential
processes are those that might be omitted, such as planning, estimating and
staging meetings. In particular, he discusses how practices are routinely
removed: ``With most methodologies and approaches, very few stick to the
letter of them and they are always adapted, so we adapted ours to the way we
wanted it to work for us, for our own size and scale''~\cite{Coleman2008}.

Coleman's network is characterized by the ``cost of process'' (core
category) and all the factors that in management contributed to the lack of
software process improvements (SPI). The cost of process represents the
lack of formal and prescriptive work-flows in development, mainly conducted by
verbal communication without heavy documentation or bureaucracy. Coleman 
reports on the practitioners' perception that documentation alone does not 
ensure a shared understanding of project requirements. Moreover,
managers perceive rigid processes as having a negative
impact on the creativity and flexibility of the development team. This is in
accordance with our generated theory, which bases the reasons for adopting
evolutionary and low-precision engineering elements on the flexibility and
reactivity attributes of the development process in startups.

As also reported in the GSM, the definition of a ``minimum process'' is not a 
matter of poor knowledge and training, but rather a necessity that lets the 
company move faster. 
``One-size-fits-all'' solutions have always found difficulty in penetrating 
small software organizations~\cite{Staples2007}. When startups begin 
establishing any rigid SPI process, they experience process 
erosions~\cite{Coleman2008}, which result in work-flows barely satisfying 
organizational business needs. Software startups favor the use of agile 
principles in support of creativity and flexibility instead of 
one-size-fits-all SPI. 

Further, Coleman describes a management approach oriented towards
``embrace and empower'', consisting of trust in the development 
staff to carry out tasks with less direct supervision~\cite{Coleman2008}.
Nevertheless, software development managers and founders still have an impact on
management style and indirectly on the software development process. In 
early-stage startups, founders are mainly software development managers as
CEOs/CTOs and technical practitioners at the same time. As Coleman identified
the influence of the founders' and managers' background on the software
development process, the GSM similarly identifies that the CEOs/CTOs
background shapes the high-level strategies adopted in developing the initial
product. 

\begin{table*}
	\caption{GSM categories' overlap with the SMS~\cite{SMS}}
	\label{tab:an:literature_comp}
	\footnotesize
	\centering
	\begin{tabular}{lccccccccc}
		\toprule
		Author (year) & CAT1 & CAT2 & CAT3 & CAT4 & CAT5 & CAT6 & CAT7 & Count 
		& Ref. \\
		\midrule
		\midrule
		Sutton (2000) & X & X & X & X & X & X & X & 7 & \cite{Sutton2000} \\
		Kajko-Mattson (2008) & X & X & X & X & X & X & X & 7 &
		\cite{Kajko-Mattsson2008} \\
		Crowne (2002) & X & X & X & X & X & X & X & 7 & \cite{Crowne2002} \\
		Coleman (2008) & X & X & X & X & X & X & X & 7 & \cite{Coleman2008a} \\
		Coleman (2008) & X & X & X & X & X & X & X & 7 & \cite{Coleman2008} \\
		Coleman (2007) & X & X & X & X & X & X & X & 7 & \cite{Coleman2007} \\
		Carmel (1994) & X & X & X & X & X & X & X & 7 & \cite{Camel1994a} \\
		Yoffie (1999) & X & X & X & X & X & X & & 6 & \cite{Yoffie1999} \\
		Zettel (2001) & X & X & X & X & & & X & 5 & \cite{Zettel2001} \\
		Jansen (2008) & X & & X & & X & X & X & 5 & \cite{Jansen2008} \\
		Heitlager (2007) & & & X & X & X & X & X & 5 & \cite{Heitlager2007} \\
		Deias (2002) & X & X & & X & X & X & & 5 & \cite{Deias} \\
		Ambler (2002) & X & X & & & X & X & X & 5 & \cite{Ambler2002} \\
		Wood (2005) & X & & X & X & & & X & 4 & \cite{Wood2005} \\
		Tingling (2007) & & X & & X & X & X & & 4 & \cite{Tingling2007} \\
		Taipale (2010) & X & X & & X & X & & & 4 & \cite{Taipale2010} \\
		da Silva (2005) & X & X & & X & X & & & 4 & \cite{Silva2005} \\
		Mirel (2000) & X & & X & X & X & & & 4 & \cite{Mirel2000} \\
		Midler (2008) & X & X & & X & & & X & 4 & \cite{Midler2008} \\
		Tanabian (2005) & X & & & X & & & X & 3 & \cite{Tanabian2005} \\
		Stanfill (2007) & X & & & & X & & X & 3 & \cite{Stanfill2007} \\
		Mater (2000) & X & X & & & & X & & 3 & \cite{Mater2000} \\
		Kuvinka (2011) & X & X & & X & & & & 3 & \cite{Kuvinka2011} \\
		Deakins (2005) & X & X & & X & & & & 3 & \cite{Deakins2005} \\
		Yogendra (2002) & X & X & & & & & & 2 & \cite{Yogendra2002} \\
		Wall (2001) & X & & & & & & X & 2 & \cite{Wall2001} \\
		Su-Chuang (2007) & X & & & X & & & & 2 & \cite{Su-Chan2007} \\
		Steenhuis (2008) & & & & X & & & X & 2 & \cite{Steenhuis2008} \\
		Sau-ling Lai (2010) & & X & & & & & X & 2 & \cite{Lai2010} \\
		Kakati (2003) & & X & & X & & & & 2 & \cite{Kakati2003} \\
		Himola (2003) & X & & & & & & X & 2 & \cite{Hilmola2003} \\
		H\"{a}sel (2010) & X & & & X & & & & 2 & \cite{Hasel2010} \\
		Hanna (2010) & X & & & & & & X & 2 & \cite{Hanna2010} \\
		Bean (2005) & X & & & & X & & & 2 & \cite{Bean2005} \\
		Kim (2005) & & X & & & & & & 1 & \cite{Kim2005} \\
		Fayad (1997) & & & & X & & & & 1 & \cite{Fayad1997} \\
		Chorev (2006) & & & & X & & & & 1 & \cite{Chorev2006} \\
		\midrule
		Count & 29 & 22 & 13 & 26 & 18 & 14 & 20 & &\\
		\bottomrule
	\end{tabular}
\end{table*}

Baskerville~\cite{Internet} refers to rigid SPI approaches as
typically effective only in large-scale, long-term development efforts with
stable and disciplined processes. Internet-speed software development 
(oriented towards daily builds, aimed at developing a product with high speed) 
differs from traditional software development. Baskerville studied 10 companies 
using a Grounded Theory approach. He found that the major causal factors 
that influence development are a desperate rush to market, a new and unique 
software market environment, and a lack of experience developing software under 
the conditions this environment imposes.
Even though with different research focus and study context, Baskerville 
revealed similar causal factors as the GSM (see 
Table~\ref{tab:framework_comparison}). He argues that the dawn of the Internet 
era has intensified software development problems by emphasizing shorter cycle 
times as a strategy to efficiently validate a product to the target market. 

With a wider focus, Brooks~\cite{BrooksJr1987} discusses the 
challenges involved in constructing software products. Brooks divides
difficulties in development into essence (inherent to the nature of the
software), and accidents (difficulties attending software production that
are not inherent). In other words, essence concerns the hard part of building a
software through activities such as specification, design, testing. Accidents
refer to the labor of representing the software or testing its representation.
Brooks claims that the major effort applied by engineers was dedicated towards 
accident problems, trying to exploit new strategies to
enhance software performance, reliability and simplicity of development, such as
the introduction of high-level languages for programming. Despite the great
achievements in improving development performance, the ``essence'' property
of the software remained unaltered.
The basic mitigation strategies presented by Brooks on the essence (i.e. buy 
versus build; requirements refinement and rapid prototyping; incremental 
development; and great teams) accurately fit the GSM (see Table 
\ref{tab:framework_comparison}), forecasting the state-of-practice in 
modern startups.

\subsection{Theoretical categories and existing literature}
\label{sect:theory:validation:sms}

In this subsection we extend the theory validation by mapping the categories of 
the GSM to empirical studies that investigated startup companies. We map the 
studies' main contributions to one or more GSM categories 
(Table~\ref{tab:an:literature_comp}). We sorted the table according to the 
number of GSM categories covered by the studies.

Seven out of 37 studies address all GSM categories in their discussion. All 
studies address at least one GSM category. The majority of the retrieved 
studies (29) mention issues related to \textit{speed up development} (CAT1), 
the core category of the GSM. Another common category, addressed by 26 studies, 
is the \textit{team is the catalyst of development} (CAT4). The importance of 
people has been widely discussed in other software engineering studies (e.g. 
Cooper~\cite{Cooper1986}, DeMarco~\cite{peopleware-demarco1999}, 
Coleman~\cite{Coleman2004}, Valtanen~\cite{Valtanen2008}, Adolph and 
Kruchten~\cite{Adolph2011}, and Cockburn~\cite{people-first-order}), advocating 
for the need to empower people. 
Less than half of the studies mention results related to \textit{product 
quality has low priority} (CAT3), \textit{accumulated technical debt} (CAT5), 
and \textit{initial grow hinders performance} (CAT6). This indicates a 
potential lack of research and suggests directions for future work.

\subsection{Confounding factors} \label{sect:an:comp:cat-vs-literature:conf}
\begin{table*}[t]
	\caption{Confounding factors in the GSM}
	\label{tab:confounding}
	\centering
	\footnotesize
	\begin{tabular}{p{5cm}p{12cm}}
		\toprule
		Confounding factors & Description\\ 
		\midrule
		\midrule
		Creativity and innovation (\cite{Heitlager2007}) & The study reports 
		how product-oriented development, in contrast to  process imposition, 
		provides a degree of freedom to the development team that enhances the 
		creativity of developers and augments the innovation capability of the 
		company in the early-stage. \\ 
		\midrule
		Market requirements (\cite{Coleman2007}) and application type 
		(\cite{Sutton2000,Coleman2008,Coleman2008a}) & Their main impact is 
		related to the adoption of flexible and reactive  solutions for the 
		development process. In particular, the studies refer to the  necessity 
		of fulfillment of quality concerns that goes beyond scalability and UX, 
		when requirements are rigidly imposed or the application domain is 
		well-known. In these cases providing low-quality products 
		to final users might determine the failure of a startup. \\ 
		\midrule
		Developer experience (\cite{Crowne2002, Yoffie1999}) & Startups often 
		rely at the beginning on clever, but inexperienced developers. However, 
		having team members with deep experience would be a ``double-edged 
		sword''. Experience might quickly provide structure and 
		maturity to the development process; yet it might cause challenges in 
		managing self-confident overachievers that almost 
		inevitably clash. Consequently, team management might require control 
		and coordination activities that hinder flexibility of the development 
		environment which is essential in early-stage startups.\\
		\bottomrule
	\end{tabular}
\end{table*}

The purpose of this subsection is to identify which confounding factors might 
threaten the validity of the GSM. While the mapping in 
subsection~\ref{sect:theory:validation:sms} validated the literature coverage 
of GSM's categories, here we are interested in those variables that are 
\emph{not} covered by the GSM and might interfere with the theoretical model 
positively or negatively~\cite{ColinRobson2009}. We report those factors 
identified by the SMS, but not considered by the GSM: creativity 
and innovation, market requirements and application type, and developer 
experience, summarized in Table~\ref{tab:confounding}.

Understanding the impact of a confounding factor on the interpretation of the 
model is important for further analyses and use of the GSM. A researcher, 
using the GSM (Section~\ref{res:gsm}) and its implications 
(Section~\ref{sect:theory:impl}), has to contextualize his analysis with the 
startups' basic demographic and background characteristics. For example, market 
requirements (see Table~\ref{tab:confounding}) might undermine the 
generalizability of the GSM. In such a scenario, avoiding minimum expectations 
of quality assurance in ``quality critical markets'', such as security in 
banking services, would profoundly affect the customers' satisfaction.

\section{Threats to Validity} \label{valt} 

In this section we discuss the validity of the overall research methodology. 
We structure the discussion according to Wohlin's taxonomy~\cite{Wohlin2000}.

\subsection{External validity}

One threat to external validity is the selection of subjects interviewed for
the study. This threat affects GT, a qualitative research method 
using semi-structured interviews, and centered on respondent's opinions. To 
mitigate this threat we selected interviewees that covered the 
positions of CTOs and CEOs. Their broad perspectives on their startup 
organization was the only data taken into consideration in the study.

The majority of the studied startups are successful web companies, introducing 
a potential bias in the development of the GSM. In particular, we lack the 
perspective of failed startups that potentially could have provided stronger 
evidence for the relationships in the GSM. We partially mitigated this threat 
by comparing the GSM with similar models. 
The comparison helped in establishing the context to which the study findings 
can be generalized. In particular the previous model developed by
Coleman~\cite{Coleman2008} has allowed us to identify similarities and 
differences, enabling a broader reasoning related to factors that hinder 
maturing processes in startups. In addition, we analyzed literature covered by 
the SMS on startups. However, including companies focusing on e.g. 
embedded real time systems or failed startups might have led to 
different results. 

\subsection{Internal validity} 
To enhance internal validity, we created a three-dimensional research 
framework. Through a Grounded Theory approach, supported by a 
systematic mapping study, interviews and follow-up questionnaires, we searched 
for convergence among different sources of information to confirm or contradict 
the generated theory. Our strategy included also the collection of supporting 
artifacts (e.g. project plans, meeting notes, bug repositories) to verify the 
statements made by the interviews. However, none of the companies could provide 
access to this information. Furthermore, the only a subset (9 out of 13) of the 
interviewees returned the questionnaire.

To validate the GSM we conducted a comparison of the emergent theory with
existing literature and previously developed models. With the
theory validation we highlighted and examined similarities, contrasts and
explanations~\cite{Eisenhardt2007}. In this regard, Eisenhardt
stated: ``Tying the emergent theory to existing literature enhances the internal
validity, generalizability, and theoretical level of the theory building from a
case study research [\ldots] because the findings often rest on a very limited
number of cases.'' We identified important confounding factors, related to 
innovation, market requirements and developer experience (see 
subsection~\ref{sect:an:comp:cat-vs-literature:conf}). These factors are not 
catered for in the GSM, even though they are regarded (by other studies) to be 
relevant for the startup context. 

We mitigated reporting bias by packaging all needed material for
conducting new studies, providing an interview package with instructions 
available in the supplemental material of this 
paper~\cite{giardino_supplementary_2015}. Moreover, two researchers 
not involved in the execution of the study conducted a peer-review analysis of 
the theory's constructs. To control distortion during analysis we made 
extensive use of memos and comparative analysis, through which we were able to 
check if data fit into the emerging theory and countered subjectivity.

\subsection{Construct validity} \label{construct_validity}
One threat to this study is a possible inadequate description of constructs. To 
diminish this risk, the entire study constructs have been adapted to 
the terminology utilized by practitioners and defined at an adequate level for 
each theoretical conceptualization. For instance, we defined \textit{Time 
shortage} in terms of \textit{Investor pressure, CEO/business pressure, Demo 
presentations at events and internal final deadline} as used by most of the 
interviewees in the study.
Moreover, during the coding of interview transcripts, we adopted explanatory
descriptive labels for theoretical categories, to capture the underlying
phenomenon without losing relevant details. 

The second important threat is caused by the fact that interviewees might 
already be aware of the possible emergent theories analyzed by researchers. To 
reduce this risk, we did not disclose any goal or emergent results to the 
interviewees. 

\subsection{Conclusion validity}

Grounded Theory has been applied by other researchers in similar
contexts to attest relationships among conceptualizations of an examined
phenomenon (see~\cite{Coleman2007, Basri, Coleman2008a}). Those relationships
should be verified in such a way that emerging findings remain consistent as 
further data is collected. In particular we were prepared to modify generated 
categories so that the new data could be adapted into the emerging theory 
according to the concepts of theoretical sampling and saturation.

According to the theoretical sampling concept, we adjusted our study design
and the emergent theory until only marginal results were generated. Moreover, to
enhance reliability of the outcome conceptualizations and relations, we 
conducted the coding of interviews by following a systematic process. 

An important issue is related to the fact that the limited number of interviews
might not represent the complete scenarios in our study context. This issue is 
partially mitigated as result of the theoretical saturation concept. 
Ramer~\cite{Ramer}, comparing quantitative to qualitative studies, states that: 
``reaching data saturation, which involves obtaining data until no new 
information emerges, is critical for obtaining applicability in qualitative 
research''. After attesting that no more relevant information could be gathered 
from executing additional interviews, we iterated the Grounded Theory cycle one 
more time, verifying that the explanatory power of the core category was 
fulfilled.


\section{Conclusion} \label{conc} 
Startups are able to produce cutting-edge software products with a wide impact 
on the market, significantly contributing to the global economy. Software 
development, especially in the early-stages, is at the core of the companies' 
daily activities. Despite their high failure rate, an earlier systematic 
mapping study~\cite{SMS} found that the proliferation of startups 
is not matched by a scientific body of knowledge. 
To be able to intervene on software development strategies of startups with 
scientific and engineering approaches, the first step is to understand 
startups' behavior. 
Hence, in this paper, we provide an initial explanation of the 
underlying phenomenon by means of a Grounded Theory study based on 13 cases. We 
focused on early engineering activities, from idea conception to the first 
open beta release of the product.

We grounded the Greenfield Startup Model (GSM) on the hindsight knowledge 
collected from practitioners with the aim of explaining how development 
strategies are engineered and practices are utilized in startups. The 
explanatory capability and correctness of the model
has been validated through systematic comparisons with the state-of-the-art. 
The SMS revealed a multi-faceted state-of-the-art, lacking support for software 
development activities in startup companies. On the other 
hand, the study presented in this paper, provides a broad set of empirical 
evidence obtained by a Grounded Theory approach. 

The overall results of this study found that the driving characteristics of 
startups were uncertainty, lack of resources, and time-pressure.
These factors influence the software development to an extent that transforms 
every decision related to the development strategies into a difficult trade-off 
for the company. 
Moreover, although startups share characteristics with similar 
SE contexts (e.g. market-driven development, small companies and web 
engineering), a unique combination of factors poses a whole new set of 
challenges that need to be addressed by further research. 
When bringing the first product to market, startups' most urgent 
priority is releasing the product as quickly as
possible to verify the product/market fit, and to adjust the business and
product trajectory according to early feedback and collected metrics. At this
stage, startups often discard formal project management, documentation,
analysis, planning, testing and other traditional process activities.
Practitioners take advantage of an evolutionary prototyping approach, using
well-integrated tools and externalizing complexity to third party solutions.

However, the need to restructure the product and control the engineering 
activities when the company grows counterbalances the initial gain of 
flexibility and speed. If successful, the startup will face growth of 
customers, employees and product functionalities that leads to the necessity of 
controlling the initial chaotic software development environment. The most 
significant challenge for early-stage startups is finding the sweet spot 
between being fast enough to enter the market early and controlling the amount 
of accumulated technical debt. 

What follows from the GSM are four software development objectives that need to 
be considered by early-stage startups and researchers seeking to improve 
state-of-the-art:
\begin{compactitem}
\item Integration of scalable solutions with fast iterations and minimal
set of functionalities.
\item Empowerment of the team-members granting them the responsibility and 
autonomy to be involved in all activities of the development phase.
\item Improvement of workflow patterns through the initiation of a minimal 
project management.
\item Adaptation of Lean and Agile development practices after the initial 
chaotic startup phase.
\end{compactitem}

In this paper we discussed a number of novel challenges for both
practitioners and researchers, while presenting a first set of concepts, terms
and activities for the rapidly increasing startup phenomenon. By making a 
comparison with Berry's definition of SE~\cite{Berry1992}, we would like to see 
the rise of a new discipline - \textit{startup engineering} - which can be 
defined as \textit{the use of scientific, engineering, managerial and 
systematic approaches with the aim of successfully developing software systems 
in startup companies}.


%

\ifCLASSOPTIONcompsoc
  \section*{Acknowledgments}
\else
  \section*{Acknowledgment}
\fi

The authors would like to thank the Blekinge Institute of Technology (Sweden)
and the Free University of Bolzano (Italy), all the participants for their
support of this research, and Philip Stastny for proofreading the manuscript.

\ifCLASSOPTIONcaptionsoff
  \newpage
\fi







\begin{IEEEbiography}[{\includegraphics[width=1in,height=1.25in,clip,keepaspectratio]
{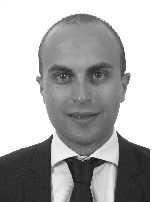}}] {Carmine Giardino} received a BSc degree in Computer Science 
from the University of Salerno in 2010, a MSc in Software Engineering at the 
Free University of Bolzano/Bozen and Blekinge Institute of Technology in 2013. 
He is a PhD student at the Free University of Bolzano/Bozen. His research 
interests include software startups and information services with focus on 
trading securities. Contact him at carmine.giardino@gmail.com
\end{IEEEbiography}

\begin{IEEEbiography}[{\includegraphics[width=1in,height=1.25in,clip,keepaspectratio]
{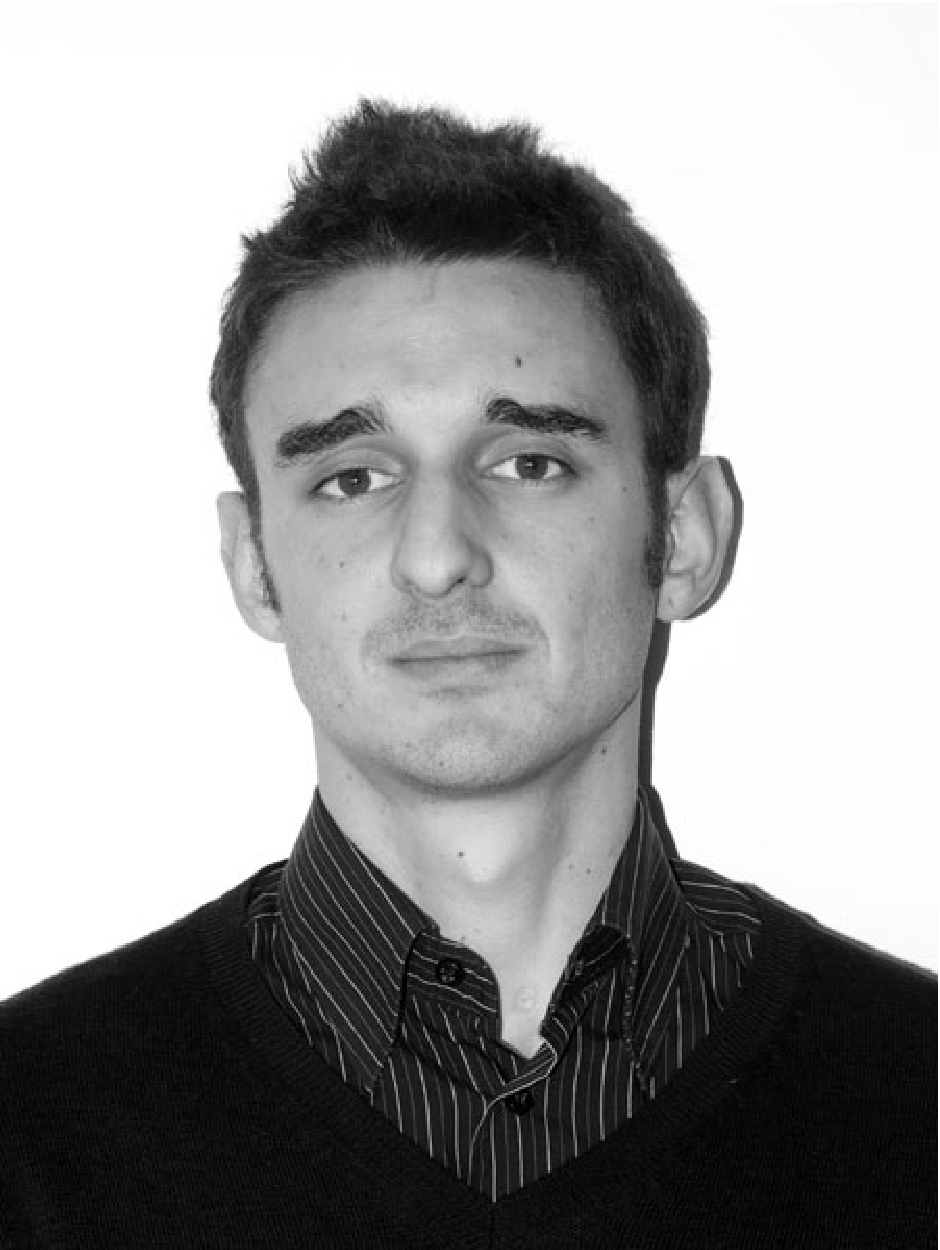}}]{Nicol\'o Paternoster} received a BSc degree in Applied 
Mathematics from the University of Roma - Tor Vergata in 2009 and a MSc in 
Software Engineering at the Free University of Bolzano/Bozen and Blekinge 
Institute of Technology in 2013. He works as freelance software engineer and 
consultant mainly for early-stage startups. His research interest includes 
software startups and blockchain technology. For more information or contact: 
http://adva.io 
\end{IEEEbiography}

\begin{IEEEbiography}[{\includegraphics[width=1in,height=1.25in,clip,keepaspectratio]
{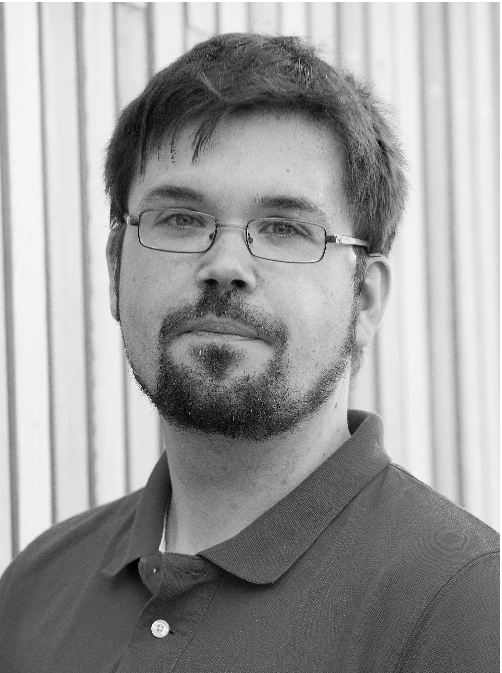}}]{Michael Unterkalmsteiner} received a BSc degree in Applied 
Computer Science from Free University of Bozen-Bolzano in 2007, a MSc 
and PhD degree in Software Engineering at Blekinge Institute of Technology 
(BTH) in 2009 and 2015 respectively. He is a postdoctoral researcher at BTH. 
His research interests include software repository mining, software measurement 
and testing, process improvement, and requirements engineering. He is a member 
of the IEEE. For more information or contact: www.lmsteiner.com
\end{IEEEbiography}

\begin{IEEEbiography}[{\includegraphics[width=1in,height=1.25in,clip,keepaspectratio]
{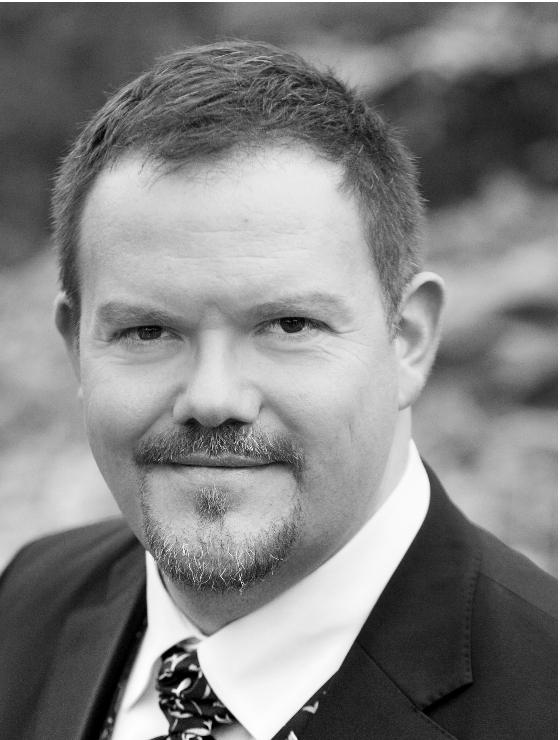}}]{Tony Gorschek} is a Professor of Software Engineering at 
Blekinge Institute of Technology (BTH). He has over ten years industrial 
experience as a CTO, senior executive consultant and engineer, but also as chief
architect and product manager. In addition he has built up five startups in
fields ranging from logistics to internet based services.
Currently he manages his own consultancy company, works as a CTO, and
serves on several boards in companies developing cutting edge technology
and products. His research interests include requirements engineering,
technology and product management, process assessment and improvement,
quality assurance, and practical innovation. www.gorschek.com
\end{IEEEbiography}

\begin{IEEEbiography}[{\includegraphics[width=1in,height=1.25in,clip,keepaspectratio]
{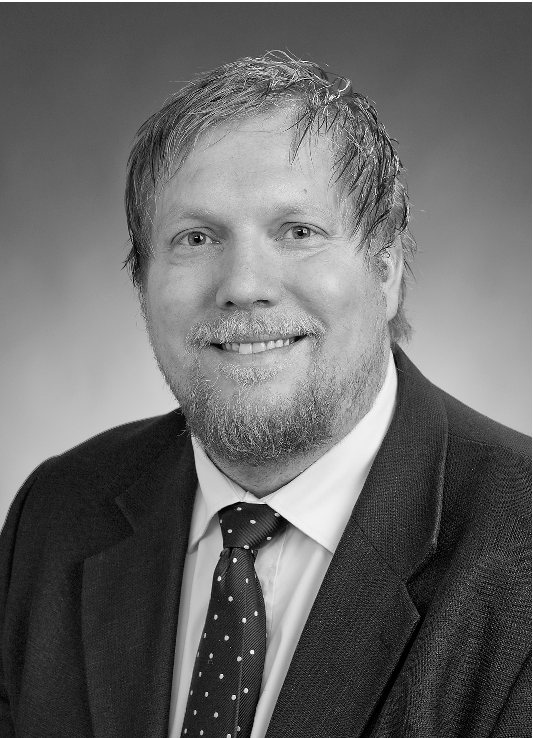}}] {Pekka Abrahamsson} received the PhD degree in software 
engineering from the University of Oulu, Finland, in 2002. He is a full 
professor of software engineering at the Department of Computer and Information 
Science, Norwegian University of Science and Technology, NTNU, 
Trondheim, Norway. Prior to his current appointment, he has served in 
professor positions at the Free University of Bozen Bolzano, University 
of Helsinki and VTT Technical Research Centre of Finland. His research 
interests are in the empirical software engineering, software startups 
and innovation. He is the recipient of Nokia Foundation Award in 2007 
for his merits in agile software development research and his European 
research project on Agile methods in embedded systems received ITEA 
Achievement Silver award in 2007. He heads today the global Software 
Startup Research Network and is a member of the IEEE and ACM.
\end{IEEEbiography}

\vfill






\end{document}